\documentclass[
 reprint,
 amsmath,amssymb,
 aps,
 prb,
]{revtex4-2}

\usepackage[utf8]{inputenc}
\usepackage{graphicx}                     
\usepackage{lmodern}                      
\usepackage{siunitx}                      
\usepackage[english]{babel}
\usepackage{hyperref}                     
\usepackage{booktabs}                     
\usepackage{physics}                      
\usepackage{ltablex}
\usepackage{xcolor}                       

\graphicspath{{figures/}}

\begin{document}

\preprint{APS/123-QED}

\title{Spread balanced Wannier functions: Robust and automatable orbital localization}

\author{Pietro F. Fontana}
 \affiliation{CAMD, Department of Physics, Technical University of Denmark, DK-2800 Kongens Lyngby, Denmark 
}
 \author{Ask H. Larsen}
 \affiliation{CAMD, Department of Physics, Technical University of Denmark, DK-2800 Kongens Lyngby, Denmark 
}
 
 \author{Thomas Olsen}
 \affiliation{CAMD, Department of Physics, Technical University of Denmark, DK-2800 Kongens Lyngby, Denmark 
}
\author{Kristian S. Thygesen}%
 \email{thygesen@fysik.dtu.dk}
\affiliation{CAMD, Department of Physics, Technical University of Denmark, DK-2800 Kongens Lyngby, Denmark 
}

\date{\today}

\begin{abstract}
We introduce a new type of Wannier functions (WFs) obtained by minimizing the conventional spread functional with a penalty term proportional to the variance of the spread distribution. This modified Wannierisation scheme is less prone to produce ineffective solutions featuring one or several poorly localized orbitals, making it well suited for complex systems or high-throughput applications. Furthermore, we propose an automatable protocol for selecting the initial guess and determine the optimal number of bands (or equivalently WFs) for the localization algorithm. The improved performance and robustness of the approach is demonstrated for a diverse set of test systems including the NV center in diamond, metal slabs with atomic adsorbates, spontaneous polarization of ferroelectrics and 30 inorganic monolayer materials comprising both metals and semiconductors. The methods are implemented in Python as part of the Atomic Simulation Environment (ASE). 
\end{abstract}

\maketitle

\section{Introduction}
As computational codes are being increasingly automated it becomes possible to perform complex, high-throughput investigations with minimal human efforts, creating new advantages and opportunities. Within materials science, these developments significantly expand the range of materials/properties that can be examined by a single researcher. Moreover, it increases data quality by reducing the risk of human errors, and enables researchers to address materials phenomena or properties outside his/her domain expertise by lowering barriers related to the technical aspects of the calculation. While some computational tasks are straightforward to automate, others are more challenging. An important example of the latter is the generation of localized representations of the delocalized Bloch states of a crystal, i.e. Wannier functions \cite{wannier} (WFs). In a seminal paper, Marzari and Vanderbilt \cite{mlwf1997} introduced a practical scheme for calculating maximally localized Wannier functions (MLWFs) that overcomes the problem of the non-uniqueness (or "gauge dependence"). For simple systems, i.e. when the bands of interest form an isolated group and/or there are few atoms in a unit cell, the standard algorithms typically yield well localized WFs. In general, however, the construction of useful WFs requires some hand-holding rendering automation highly non-trivial. 

When the bands of interest (from hereon referred to as the "target bands") are isolated from all higher and lower lying bands by energy gaps, the MLWFs are obtained by minimizing the sum of the quadratic spread of all the WFs. In the general case, i.e. when the target bands do not form an isolated group, the problem of finding a proper localized representation becomes significantly harder. In this case, extra degrees of freedom (EDF) in the form of states orthogonal to the target bands, must be included to aid the localization. Since the target bands typically contain the occupied manifold, the problem of identifying the optimum EDF can be seen as the process of augmenting the target bands by their anti-bonding states \cite{prl}.      

The so-called disentanglement procedure \cite{mlwf2001} identifies the EDF by minimizing the dispersion of the $\mathbf k$-subspaces (the subspace spanned by the target bands and the EDF at a given $\mathbf k$) across the Brillouin zone (BZ). Having identified the optimal $\mathbf k$-subspaces the MLWFs are obtained following the usual localization procedure. For perfect crystals, this approach is well suited. On the other hand, for systems where crystal momentum is not a good quantum number, e.g. molecules, amorphous solids, crystals with defects, etc. the idea of minimizing the $\mathbf k$-dispersion does not seem a natural strategy. As an alternative, one can determine the EDF by direct minimization of the spread functional. With this strategy, the selection of the EDF and the localization into WFs is cast as one global optimization problem rather the two-step strategy applied by the disentanglement procedure. This idea leads to the partly occupied Wannier functions (POWF) developed in Refs. \cite{prl,prb}. Being the result of a one- rather than two-step optimization, the POWFs have smaller spread than the MLWFs. Moreover, the scheme avoids reference to the $\mathbf k$-dispersion, which seems natural for non-periodic systems. We note that the POWFs were rediscovered in a different but equivalent form in Ref. \cite{variational_wan}. 

Regardless of how the EDF are selected the standard localization procedure, i.e. the minimization of the sum of the quadratic spreads, is not always straightforward. One manifestation of this problem is its starting-point dependence, i.e. that different WFs are obtained depending on the initial guess for the orbitals, e.g. orbital type ($s$,$p$,$d$) and position (atom- or bond centered). This indicates that the conventional spread functional exhibits several local minima. From a practical point of view, the lack of robustness/reproducibility arising from the starting point dependence is not a problem in itself as long as a decent set of WFs is obtained. Unfortunately, even that can be challenging and sometimes requires tuning of the initial guess, the target bands, and number of EDF. One specific problem sometimes encountered is that all WFs become well localized except for one or a few which remain delocalized; this renders the entire set of WFs useless for many purposes and significantly complicates automatization. 

In this paper, we introduce a new class of spread functionals that explicitly penalize delocalization of individual WFs. The minimization of these functionals generally produces WFs with a more balanced spread distribution. In particular, the problem of "sacrificing" one WF to improve the total spread does not occur. We also introduce a specific protocol for automatically initializing the WFs based on the valence configuration of the involved atoms. Leveraging the properties of the POWFs, we device a simple trial-and-error, yet easy to automate, procedure for selecting the optimal number of EDF. Putting it all together we arrive at a highly robust and fully automatic scheme for constructing spread balanced WFs for general types of materials. We demonstrate the method for a number of challenging systems, including atoms adsorbed on metal slabs, the NV defect in diamond, and a set of 30 two-dimensional (2D) materials arbitrarily selected from the Computational 2D Materials Database (C2DB) \cite{c2db}. The WFs are used to obtain electronic band structures and spontaneous polarisations within the framework of the modern theory of polarization \cite{King-Smith1993}. All methods are implemented in the open source Atomic Simulation Environment (ASE) \cite{ase}.

\section{Theory}

In this section we review the theory and construction of partly occupied WFs. We then introduce our new spread functionals designed to produce WFs with narrow size distributions. While we have investigated several different functionals, we focus on the best performing one, the minimal variance spread functional, throughout this paper. Finally, we describe our protocols for initializing WFs and selecting the optimal number of WFs, respectively.      

\subsection{Partly occupied Wannier functions}
\label{sec:powf}

The partly occupied Wannier functions were introduced in 2005 \cite{prl, prb} and were recently demonstrated \cite{variational_wan} to represent the global minimum of the quadratic spread functional. The POWFs are related to the maximally localized Wannier functions \cite{mlwf1997, mlwf2001} but avoid explicit reference to the wave vector in the band disentanglement procedure, and are directly applicable to non-periodic systems. Instead of maximizing the reciprocal space smoothness the POWF method is entirely based on the minimization of the real space spread of the WFs.

For systems with periodic boundary conditions and a sufficiently large supercell, the minimization of the conventional Marzari-Vanderbilt spread functional \cite{mlwf1997} for a set of WFs $\{w_n(\vb{r})\}_{n=1}^{N_w}$ is equivalent \cite{resta} to the maximization of
\begin{equation}
  \label{eq:ase-fun}
  \Omega = \sum_{n=1}^{N_w} \sum_{\alpha=1}^{N_G} W_{\alpha} |Z_{\alpha, nn}|^2
\end{equation}  
where the matrix $Z_{\alpha}$ is defined as
\begin{equation}
  \label{eq:z-mat}
  Z_{\alpha, nm} = \langle w_n | e^{-i \vb{G}_{\alpha} \cdot \vb{r}} | w_m \rangle.
\end{equation}
The $\{\vb{G}_{\alpha}\}$ is a set of $N_G$ reciprocal lattice vectors that connect each $k$-point to its neighbors and $W_{\alpha}$ are corresponding weights accounting for the shape of the unit cell. The value of $N_G$ can range from 3 to 6 depending on the  symmetry of the unit cell. For a discussion about these vectors and weights we refer to Ref. \cite{berghold, silvestrelli}. 
The definition of localization we impose with the functional $\Omega$, as in the case of the Marzari-Vanderbilt spread functional, is equivalent to the Foster-Boys method \cite{foster-boys}.

We emphasize that the assumption of a large supercell with periodic boundary conditions does not represent a limitation. For example, it applies to a pristine periodic crystal (with the primitive cell repeated a number of times in all directions), isolated entities like molecules or clusters surrounded by a sufficiently large vacuum region, a surface slab (possibly with the primitive cell repeated in the in-plane directions), or a solid with periodically repeated disorder, e.g. an impurity or point defect.

The goal of this approach is to obtain a set of $N_w$ localized WFs that can reproduce any eigenstate below an energy threshold, $E_0$, exactly. Given $N_b$ available eigenstates, a localization subspace is defined as the space spanned by the $M$ eigenstates with energy below $E_0$ and additional $L$ extra degrees of freedom (EDF), where $M + L = N_w \leq N_b$.
Each WF is then defined as
\begin{equation}
  w_n = \sum_{m=1}^{M} U_{mn} \psi_m + \sum_{l=1}^L U_{M+l,n} \phi_l
\end{equation}
where the EDFs $\phi_l$ are defined as
\begin{equation}
  \phi_l = \sum_{m=1}^{N_b - M} c_{ml} \psi_{M+m}.
\end{equation}
The matrix $c$ has orthonormal columns while the matrix $U$ is unitary.

All the expressions in this section refer to the simple case where the eigenstates have been obtained in a large supercell for which a $\Gamma$-point sampling of the Brillouin zone (BZ) is a good approximation. We stress, however, that for systems exhibiting periodicity on a smaller scale than the supercell dimensions, e.g. for a perfect crystal, it is possible to formulate the theory in terms of the eigenstates of the primitive unit cell sampled on a uniform grid of $k$ points.\cite{prb} In this case, the number of fixed states, $M^{\vb{k}}$, and EDF, $L^{\vb{k}}$, become $\vb{k}$-dependent.  

The localization functional $\Omega$ can be maximized with respect to $U$ and $c$ using any gradient-dependent algorithm under the constraint of orthonormality of the EDFs, $\phi_l$, implemented e.g. via the method of Lagrange multipliers. This step is referred to as the iterative localization procedure.
As noted above, the maximization of the functional $\Omega$ is equivalent to the minimization of the Marzari-Vanderbilt spread functional \cite{mlwf1997}. 
The latter often appears in the literature as $\Omega$ but we nevertheless decided to use this symbol in order to keep the notation consistent with the previous works on POWF \cite{prl, prb}.

The POWF method was originally implemented in the open source ASE \cite{ase} Python package. All the methods described in the following were implemented as extensions/improvements to the existing POWF-ASE code. 

\subsection{Variance reducing spread functionals}
\label{sec:functionals}

The maximization of $\Omega$ in Eq. (\ref{eq:ase-fun}) is equivalent to the minimization of a cost functional given by the sum of the quadratic spreads (second moments) of the individual WFs. In our experience, this approach is not robust and can produce delocalized WFs, in particular, for large numbers of WFs ($> 50$). We hypothetize that this happens because the cost of delocalizing a single function can be compensated by a small improvement in the  localization of a number of other functions.  

\begin{figure}[t]
    \centering
    \includegraphics[width=\columnwidth]{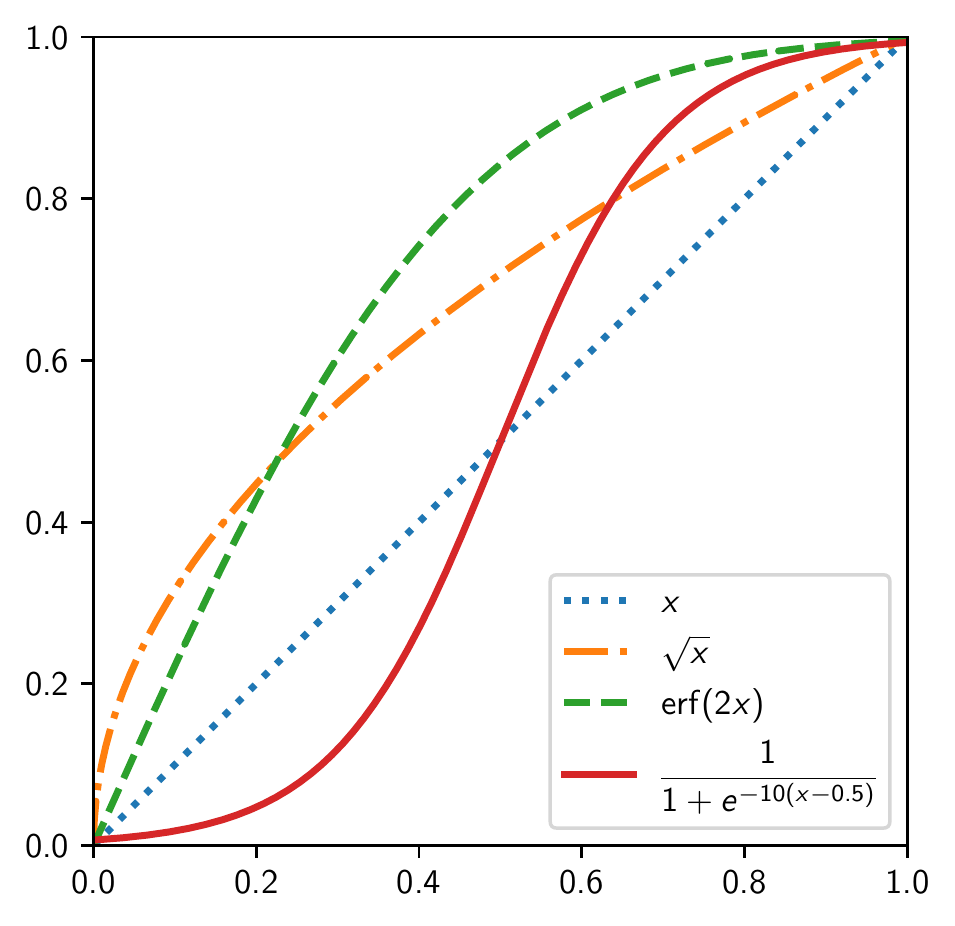}
    \caption{
      Different functions applied to the $|Z|^2$ matrix elements entering the localization functional Eq. (\ref{eq:wrap-fun}).
      The original localization functional is corresponds to $f(x)=x$ (blue dots).
    }
    \label{fig:funs}
\end{figure}

To circumvent this problem, we have explored different types of cost functionals designed to share the spread more evenly across the entire set of WFs. One approach is to apply a function on top of the $Z$-matrix elements of the original functional (these are related to the inverse spread of the corresponding WF)
\begin{equation}
  \label{eq:wrap-fun}
  \Omega_f = \sum_{n=1}^{N_w} \sum_{\alpha=1}^{N_G} W_{\alpha} f \left( |Z_{\alpha, nn}|^2 \right).
\end{equation}
We have tested different functions $f(x)$: square root ($\sqrt{x}$), scaled error function ($\frac{2}{\sqrt{\pi}} \int_0^{2x} e^{-t^2} dt$), scaled and translated sigmoid function ($1 / \left[1 + e^{-10(x-0.5)} \right]$), see Figure \ref{fig:funs}.
They span the same range as the original matrix elements, but introduce flat plateaus for well localized and a steeper slope for delocalized WFs. The effect of the steeper slope increases the gain of localizing a delocalized WF relative to the cost of delocalizing an already localized WF.
In particular, the sigmoid function has an additional penalty for delocalized functions ($|Z_{\alpha, nn}|^2 < 0.5$; the threshold can be tuned if needed). 
We mention that the modification function, $f$, may alternatively be applied to the $\alpha$-sum instead of the individual $Z$-matrix elements, but this was not pursued in the current study.

\begin{figure}[b]
    \centering
    \includegraphics[width=\columnwidth]{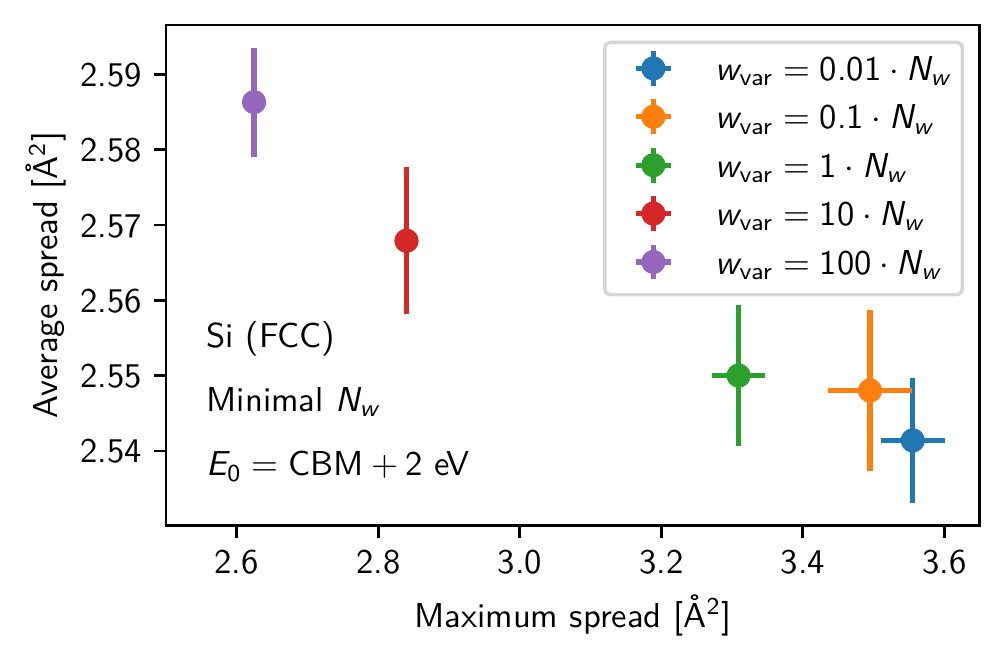}
    \caption{
      Effect of changing the weight parameter, $w_{\mathrm{var}}$, of the variance penalty term in $\Omega_{\mathrm{var}}$ for Wannier functions of bulk silicon. For larger values of $w_{\mathrm{var}}$, the spread of the most delocalised WF (maximal spread) decreases while the average spread of the WFs increases slightly. The symbols represent the mean values over 10 independent Wannierisations with different initial guess and the bars indicate the standard deviations.
      The definition of the minimal $N_w$ is given in Eq. \ref{eq:min_nw}.
    }
    \label{fig:penalty_const}
\end{figure}

Our second approach adds a penalty term to the original functional proportional to the variance of the spread distribution 
\begin{equation}\label{eq:Omega_var}
          \mathrm{\Omega_{var}} = \Omega - w_{\mathrm{var}} \mathrm{Var} \left[ \sum_{\alpha=1}^{N_G} W_{\alpha} | Z_{\alpha,nn} |^2 \right]
\end{equation}
where $w_{\mathrm{var}}$ is a parameter setting the weight for the variance term. In all our calculations we have set $w_{\mathrm{var}} = N_w$ (the weight of the penalty term should grow linearly with $N_w$ as the same holds for $\Omega$).
Of all the functionals tested in this study, $\mathrm{\Omega_{var}}$ has the best overall performance, and we therefore focus on this functional in the rest of the paper. 
In particular, the functional based on the square root only led to minor and inconsistent improvements in the localization compared to $\Omega$.
The functionals based on the error function and the sigmoid greatly increased the average spread $\bar{s}$ in order to minimize the maximum spread $s_{\mathrm{max}}$ of the set of WFs, see App. \ref{app:spread} for the definition of $\bar{s}$ and $s_{\mathrm{max}}$.
We stress that $\mathrm{\Omega_{var}}$ properly converges to real-valued WFs, as expected when reaching the global maximum \cite{mlwf-review}.

We now return to $\Omega_{\mathrm{var}}$ and the role of the weight parameter of the penalty term, $w_{\mathrm{var}}$.
Fig. \ref{fig:penalty_const} shows the average and maximum spread of the set of WFs of bulk silicon obtained by maximizing $\Omega_{\mathrm{var}}$ with different prefactors included in $w_{\mathrm{var}}$. As expected, increasing the weight of the penalty term leads to a more narrow spread distribution and thus a smaller spread of the least localised WF (maximum spread), but at the same time leads to an increase of the average spread of the WFs.
Therefore, this parameter can be tuned as needed and may even be optimized for specific applications.
The initial value of $w_{\mathrm{var}} = N_w$ worked consistently across the set of materials we tested, hence we did not perform any further optimization.

The appearance of the penalty term in the functional Eq. \ref{eq:Omega_var} implies that the mean of the spread distribution (the quantity minimized by the standard MLWFs) is balanced with the variance of the spread distribution.
For this reason we refer to the WFs resulting from the maximization of the functional $\mathrm{\Omega_{var}}$ as \emph{spread balanced WFs}.

Finally, we note that the use of different objective functions may in general lead to a different characters of the localized orbitals, as described in Sec. IIIA of Ref. \cite{mlwf-review}. In particular, the $\Omega_{\mathrm{var}}$ functional may produce WFs that differ somewhat in shape from those of the original $\Omega$ functional. A detailed investigation of this aspect is, however, beyond the scope of the current study where we focus solely on the capability of $\Omega_{\mathrm{var}}$ to produce WFs with narrow spread distributions.

\subsection{Selecting the number of Wannier functions}
\label{sec:selecting}

\begin{figure*}[t]
  \centering
  \includegraphics[width=\textwidth]{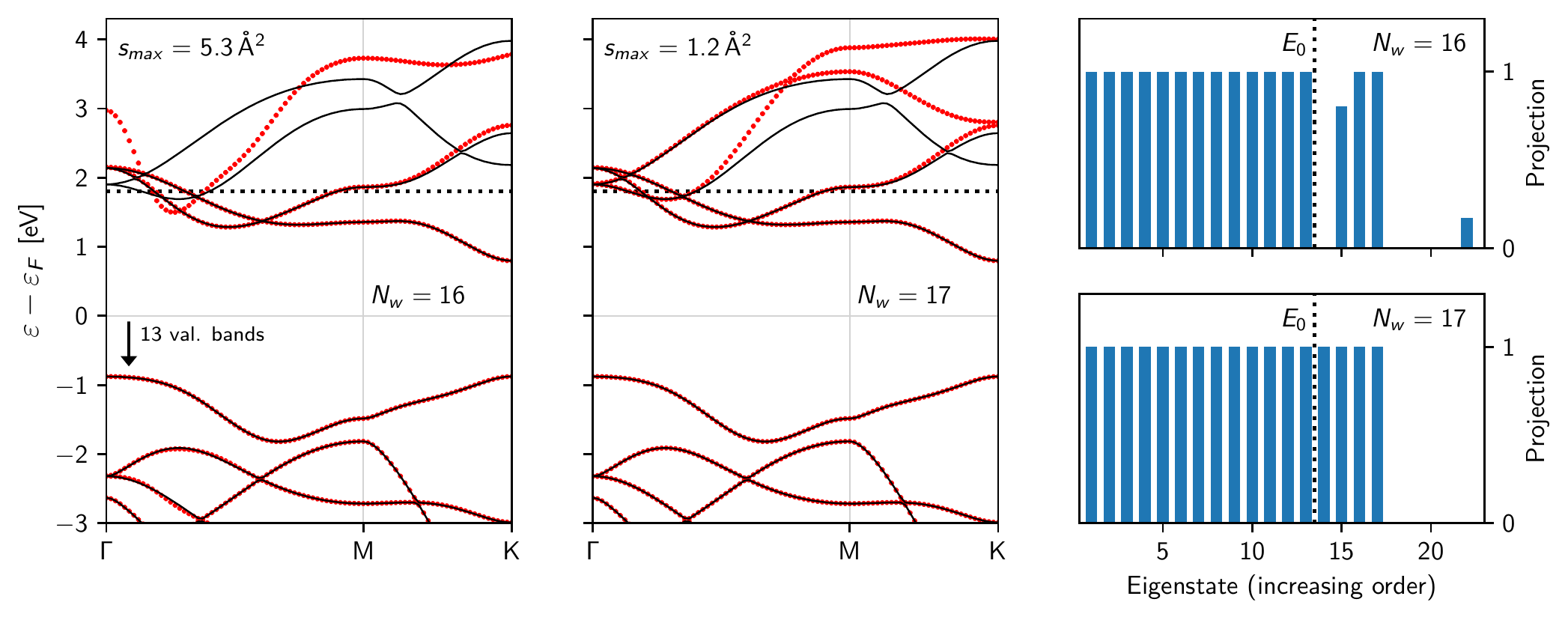}
  \caption{
    Left: 
    Interpolated band structures for monolayer MoS$_2$ (red dots) for an energy threshold  $E_0 = \mathrm{CBM} + 1$ \si{\electronvolt} (dotted black line).
    The thin black lines represent the original Kohn-Sham band structure.
    The interpolations have been performed with the WFs generated for $N_w=16$ and $N_w=17$, respectively. The former choice corresponds to the minimum number of WFs required. Inclusion of one additional WF yields better localization properties with a maximal spread of $s_{\mathrm{max}}=1.2$ \si{\angstrom^2} as compared to $s_{\mathrm{max}}=5.3$ \si{\angstrom^2}. Right:
    Projection of eigenstates on the subspace spanned by the WFs obtained for $N_w=16$ and $N_w=17$, respectively.
    The dotted black line represents the energy threshold.
    The projection is performed at the $\Gamma$ $k$-point.
  }
  \label{fig:mos2-bs}
\end{figure*}

For a given set of target bands, defined by the energy threshold $E_0$, the number of WFs, $N_w = M^{\vb{k}} + L^{\vb{k}}$, is a most important parameter for successful Wannerisation. Depending on the application, different criteria may be used to quantify the quality of a set of WFs. We have found that the spread (see App. \ref{app:spread}) of the most delocalized WF, $s_{\mathrm{max}}$, is generally a good quality-indicator, and we will use this measure along with the maximum and average band interpolation errors (see App. \ref{app:bsdist}) throughout the paper.  

In simple cases, a natural value of $N_w$ may be guessed by analysing the band structure, considering symmetries, or using chemical intuition.
In the general case, however, the optimal $N_w$ cannot be guessed and a more systematic approach is desired, see Sec. \ref{sec:optimal}.
We note that the minimum possible value for $N_w$ is given by the largest number of bands lying below $E_0$ at any $\mathbf k$, 
\begin{equation}
\label{eq:min_nw}
    N_w^{\mathrm{min}}(E_0) = \max_{\vb{k}} {\sum_n H(E_0 - \varepsilon_{n\vb{k}}) }
\end{equation}
where $H$ is the Heaviside step function.
At such $\mathbf k$-points we have $M^{\vb{k}} = N_w^{\mathrm{min}}$ and thus $L^{\vb{k}}=0$, i.e. no EDF.
In the POWF formalism there is no upper limit to $N_w$ (apart from the total number of bands available, $N_b$).

To illustrate how the $N_w$ parameter may influence the Wannierisation, we consider the case of monolayer MoS$_2$. In Fig. \ref{fig:mos2-bs}(left) we show the interpolated band structure obtained from the POWFs obtained by maximising the $\Omega$ spread functional. Results are shown with an energy threshold ($E_0$) of 1 eV above the conduction band minimum (CBM) for $N_w=16$ and $N_w=17$, respectively. (The choice $N_w=16$ corresponds to the minimum number of WFs because exactly 16 bands fall below $E_0$ at approximately 1/3 along the $\Gamma$-M path.) It is clear that the choice $N_w=17$ improves both the band structure interpolation and $s_{\mathrm{max}}$. In the present case, this result could have been anticipated by an analysis of the band structure, which present an energy gap separating the lowest 17 bands from all higher lying bands. Fig. \ref{fig:mos2-bs}(right) shows resolution of the resulting set of WF over the eigenstates (at the $\Gamma$ point). From this analysis it is clear, that the target states, consisting of states below $E_0$, can be perfectly completed by including the lowest 17 bands in the Wannierization. In contrast, a frustrated solution is obtained for $N_w=16$ where one of the EDF becomes a mix of eigenstate 15 and 22.

In general, the effect of varying $N_w$ can be difficult to predict. A prototype example where this happens is a non-elemental, low symmetry material with no natural band gaps above $E_0$. In such cases, the POWF localization algorithm, which may be seen as a bonding-antibonding completion procedure, might select EDF corresponding to high-energy eigenstates well separated from the target bands, or even mixtures of such \cite{prb}.

\subsection{Initial guess}\label{sec:initial}

The iterative localization procedure requires an initial guess for the rotation matrix $U$ and the EDF coefficient matrix $c$. The quality of the initial guess is essential. This is particularly true for systems with many WFs where the iterative optimization algorithm is more likely to get trapped in a local minimum if the initial guess is far from the global minimum.
A natural choice is to start from a set of $N_{\mathrm{AO}}$ atomic orbitals $\{ g_i\}$, and then project these onto the available eigenstates, producing the $N_b \times N_{\mathrm{AO}}$ matrix $P_{ni} = \langle \psi_n | g_i \rangle$. A prescription for extracting $U$ and $c$ from $P$ can be found in Ref. \cite{prb}. We not that $N_{\mathrm{AO}} = N_w$ in this procedure.

\section{Protocol for automated Wannierization}

In this section we present a protocol for automated construction of POWF suitable for high-throughput calculations. We stress that the protocol can be used with any spread functional, e.g. $\Omega$ or $\mathrm{\Omega_{var}}$.

\subsection{Initial guess}
\label{sec:init}

It is possible, and sometimes useful, to rely on chemical intuition when selecting a set of atomic orbitals as initial guess for the Wannierisation. On the other hand, chemical intuition is not easy to schematize in a form valid for general materials. With automation a key motivation, we therefore propose a simple, generally applicable protocol for an initial guess, which does not require any parameters and which proved to be highly effective: For each atom with \textit{d}-states in its valence electron configuration, we include a group of 5 atom-centered \textit{d} orbitals, one for each value of the magnetic quantum number. This is motivated by our observation that in such systems one typically finds WFs that closely mimick atom-centered $d$-orbitals. The total set of $d$-orbitals ($N_d$ equals 5 times the number of transition metal atoms in a unit cell) then sets a lower limit for $N_w$, in addition to the one set by the threshold energy, $E_0$. The set of $d$-orbitals is complemented by $N_s=N_w-N_d$ $s$-orbitals placed at random positions, but always within a radius of \SI{1.5}{\angstrom} of an atom. These $s$-orbitals can act as ``nucleation centers'' for atom- as well as bond-centered $s, p$ or $sp$-like WFs. 
All the aforementioned atomic orbitals are always set with a Gaussian of half-width \SI{1}{\angstrom} as radial dependence.
We did not perform any tests on materials with $f$-electrons. However, due to the highly localised nature of \textit{f}-orbitals in general, we propose to treat such states similarly to $d$-states, i.e. include a group of 7 atom-centered \textit{f} orbitals in the initial guess. 

We have found that this choice of initial guess is both effective and robust in the sense that leads to fairly rapid convergence toward solutions representing either the global minimum or a local minimum close to the global one. The variations induced by the randomness in the initial guess generally produce small variations in the resulting WFs. Nevertheless, to reduce the influence of the randomness we always perform five independent localizations and pick the best solution (see caption of Figure \ref{fig:test_set} as example).

\begin{figure}[b]
    \centering
    \includegraphics[width=\columnwidth]{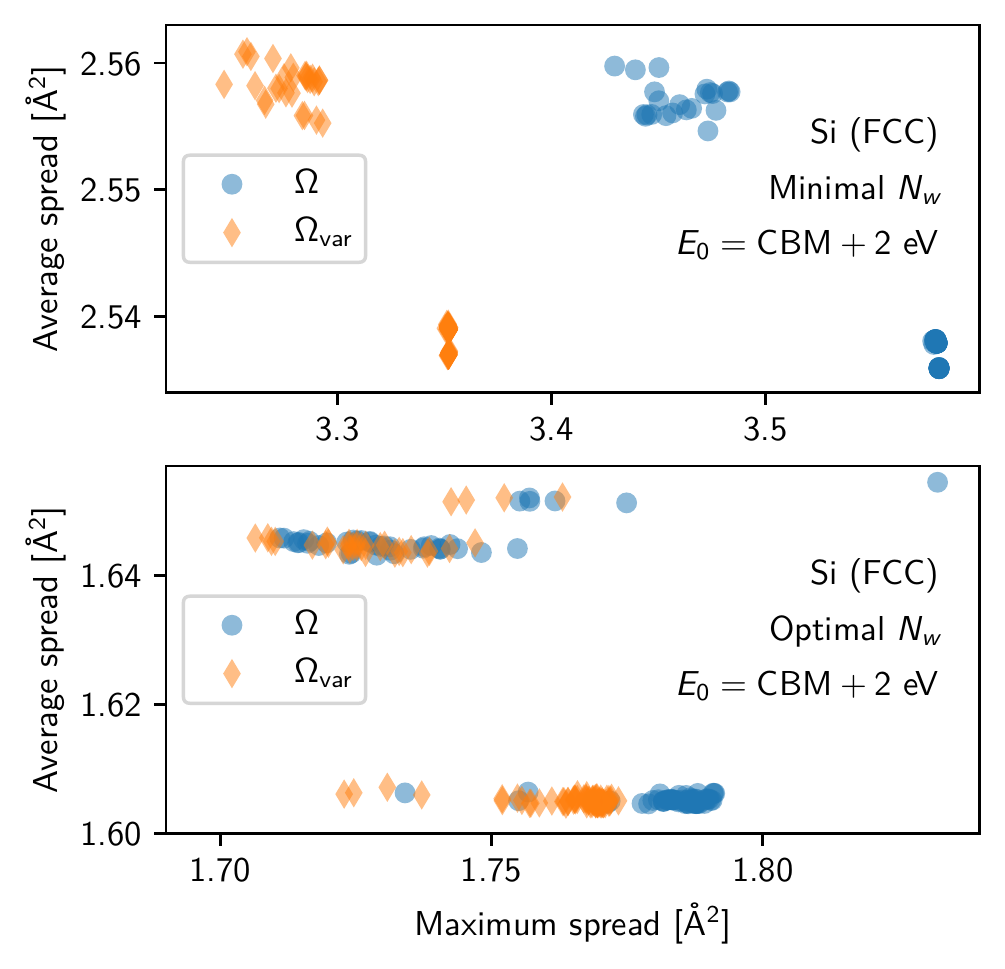}
    \caption{
      Spread distribution of several calculations with different random seeds for the initial guess.
      Each data point represents the result of a single calculation, for a total of 100 for each functional.
    }
    \label{fig:local_minima}
\end{figure}

We mention that an alternative method for initializing the Wannierisation is the selected columns of the density matrix (SCDM) method proposed by Damle et al. \cite{scdm}, that also does not require human intervention.
The SCDM did, however, not lead to improved performance over the initial guess protocol described above when applied to a set of materials with isolated groups of bands.
It is not straightforward to combine the SCDM with the disentanglement method of our POWF scheme and thus we were not able to test the SCDM for more complex situations. 

The element of randomness in our initialization scheme may seem to be a weakness. However, we see it as a strength because it allows us to repeat the Wannierisation with slightly different starting points, which often leads to slightly different outcomes of which the "best" solution can be selected. The results in Fig. \ref{fig:local_minima} show the presence of multiple nearby local maxima of both $\Omega$ and $\Omega_{\mathrm{var}}$ for bulk silicon. In this case, all the solutions are valid in the sense that the WFs are all well localized and real valued. Nonetheless, in a given situation one may prefer a specific solution satisfying certain problem specific requirements. The figure also clearly reveals the consistent reduction of the maximum spread $s_{\mathrm{max}}$ when using the $\mathrm{\Omega_{var}}$ functional, in particular when used with a non-optimal $N_w$, such as the minimum value.
The improvement becomes less pronounced in this case when using the optimal $N_w$. A more detailed discussion of these aspects are provided in Sec. \ref{sec:HT}.

\subsection{Optimal number of Wannier functions}\label{sec:optimal}

\begin{figure}[t]
    \centering
    \includegraphics[width=\columnwidth]{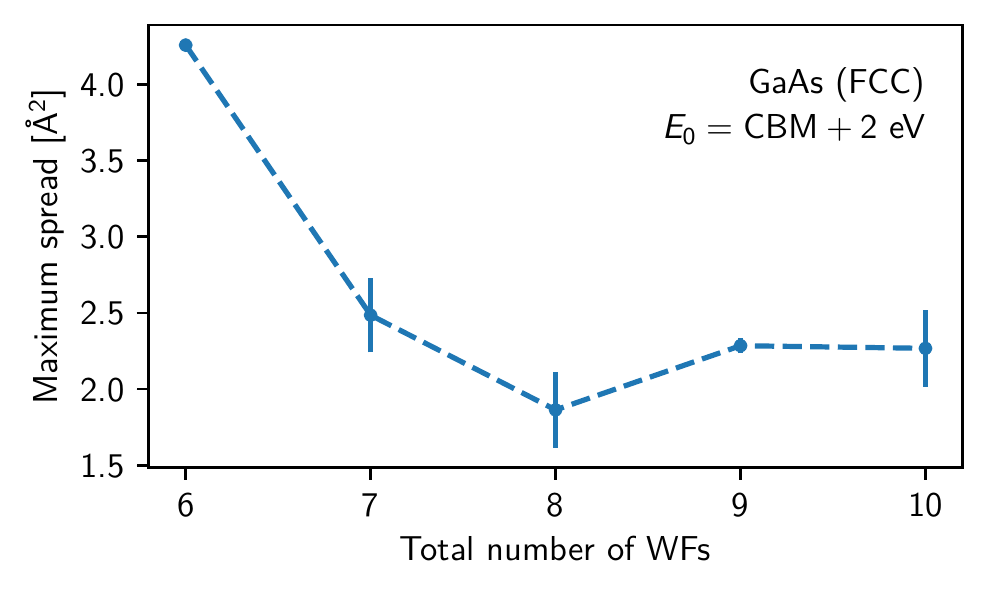}
    \includegraphics[width=\columnwidth]{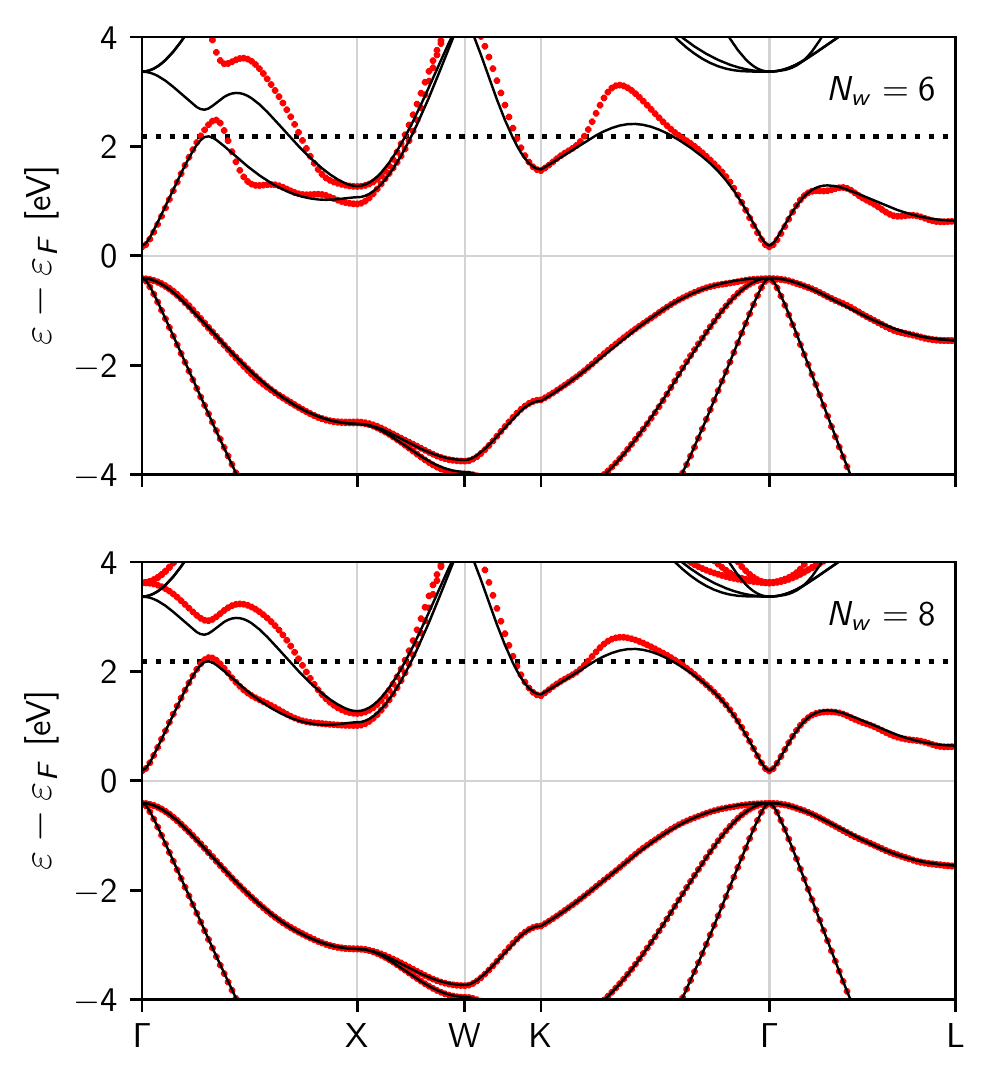}
    \caption{
      Top:
      The spread of the most delocalized WF obtained for GaAs(FCC) for different values of $N_w$.
      The calculations for each value of $N_w$ have been repeated 5 times with different initial orbitals. The circles indicate the average value and the error bars are the standard deviation over the 5 independent Wannierisations.
      Bottom:
      Interpolated band structure of GaAs obtained using the minimal $N_w = 6$ (as defined by the energy threshold of $E_0 = \mathrm{CBM} + 2$ \si{\electronvolt})  and the optimal $N_w = 8$.
      The thin black lines represent the Kohn-Sham band structure, the red dots are the Wannier interpolation.
    }
    \label{fig:gaas-best}
\end{figure}

When the optimal number of WFs cannot be guessed, c.f. Sec. \ref{sec:selecting}, it must be computed. We do this by constructing the POWFs for a range of $N_w \geq N_w^{\mathrm{min}}$ and selecting the solution presenting the smallest $s_{\mathrm{max}}$. In other words, we add EDF to the Wannierization space as long as it reduces the spread of the least localized WF. 
For the calculations presented in this work we have varied $N_w$ from  $N_w^{\mathrm{min}}$ to $N_w^{\mathrm{min}}+5$.
Depending on the size of the system (number of bands) and the available computational resources, a higher upper limit may be chosen.
Due to the randomness in the initial guess, we run 5 optimizations for each value of $N_w$ and select the best solution. In total we thus perform 25 Wannierisations for each material.
Based on our experience, the resulting optimal $N_w$ is the same for $\Omega$ and $\mathrm{\Omega_{var}}$.

As an example, Fig. \ref{fig:gaas-best} shows $s_{\mathrm{max}}$ for bulk GaAs. The vertical lines represent the variations due to the randomness in the initial guess. The minimal $\langle s_{\mathrm{max}}\rangle $ appears for $N_w = 8$. As was found for MoS$_2$, the optimal $N_w$ also produces a better interpolation of the band structure, in particular around the energy threshold ($E_0$=CBM+2 eV), see Fig. \ref{fig:gaas-best}. We do stress, however, that the size of $s_{\mathrm{max}}$ is not always directly correlated with the band interpolation error\cite{auto-wan, variational_wan}.

\section{Results}

In this section we present the results we obtained with our methods.
We start presenting the differences we observed with the new localization functional on few specific systems, then we move to a verification of the automated procedure on a set of 30 two-dimensional materials.

\subsection{An illustrative example: WMo$_3$Te$_8$}

In order to demonstrate the effect of the variance term in our newly defined localization functional $\mathrm{\Omega_{var}}$, we construct the Wannier functions of monolayer WMo$_3$Te$_8$. This material is a 2D semiconductor with 12 atoms per unit cell, 52 occupied bands, and a band gap of \SI{0.9}{\electronvolt} obtained from a DFT calculation with the PBE xc-functional \cite{pbe, c2db}.

In particular, we set an energy threshold $E_0 = \mathrm{CBM} + 2$ \si{\electronvolt} and computed 64 WFs, that is the minimal number of WFs needed to describe the states up to $E_0$.
The large number of WFs helps in proving our thesis, the optimization algorithm is in fact more likely to choose a solution with few delocalized WFs, if the relative contribution to the total localization functional is lower.

\begin{figure}[t]
    \centering
    \includegraphics[width=\columnwidth]{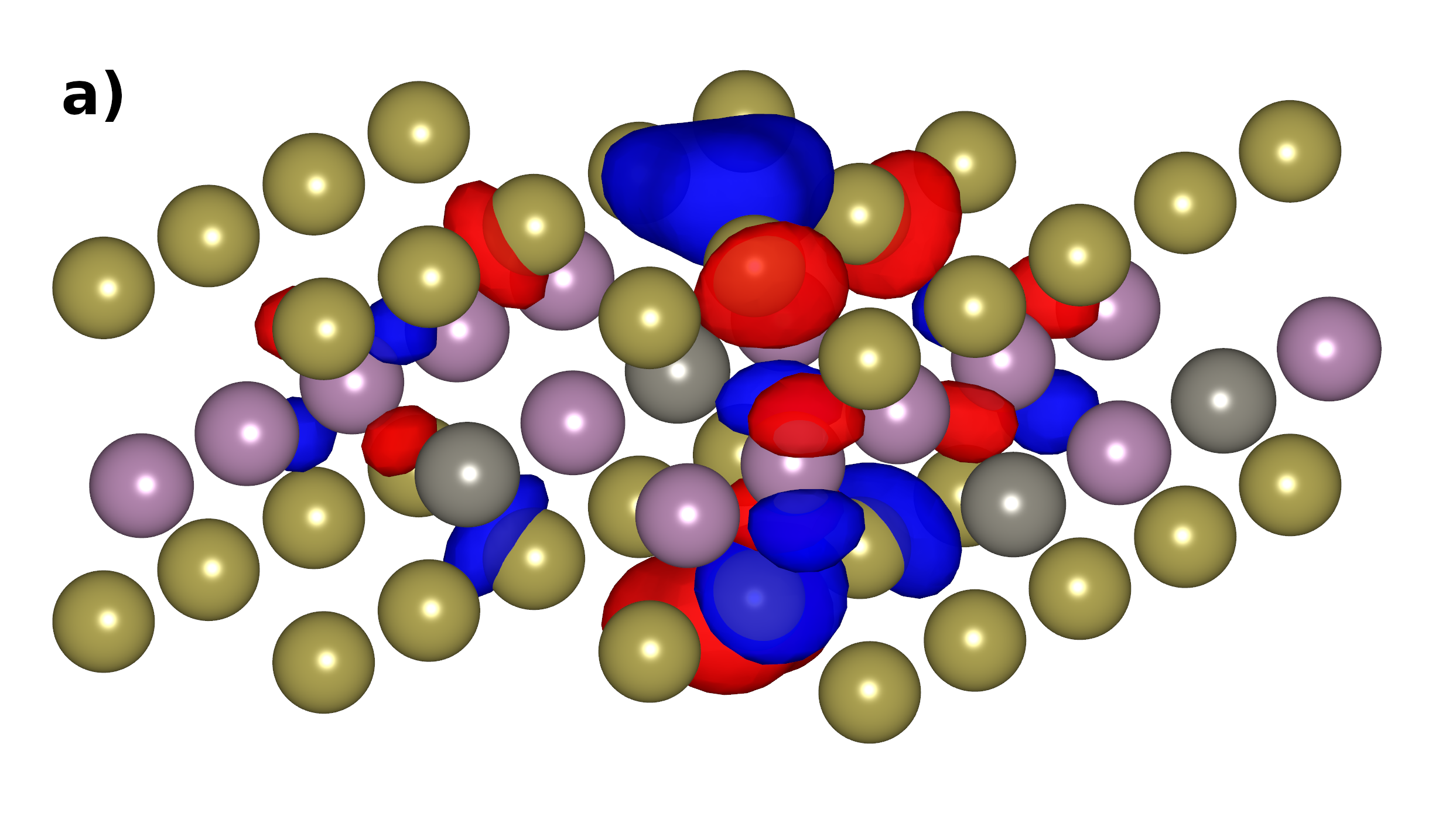}
    \includegraphics[width=\columnwidth]{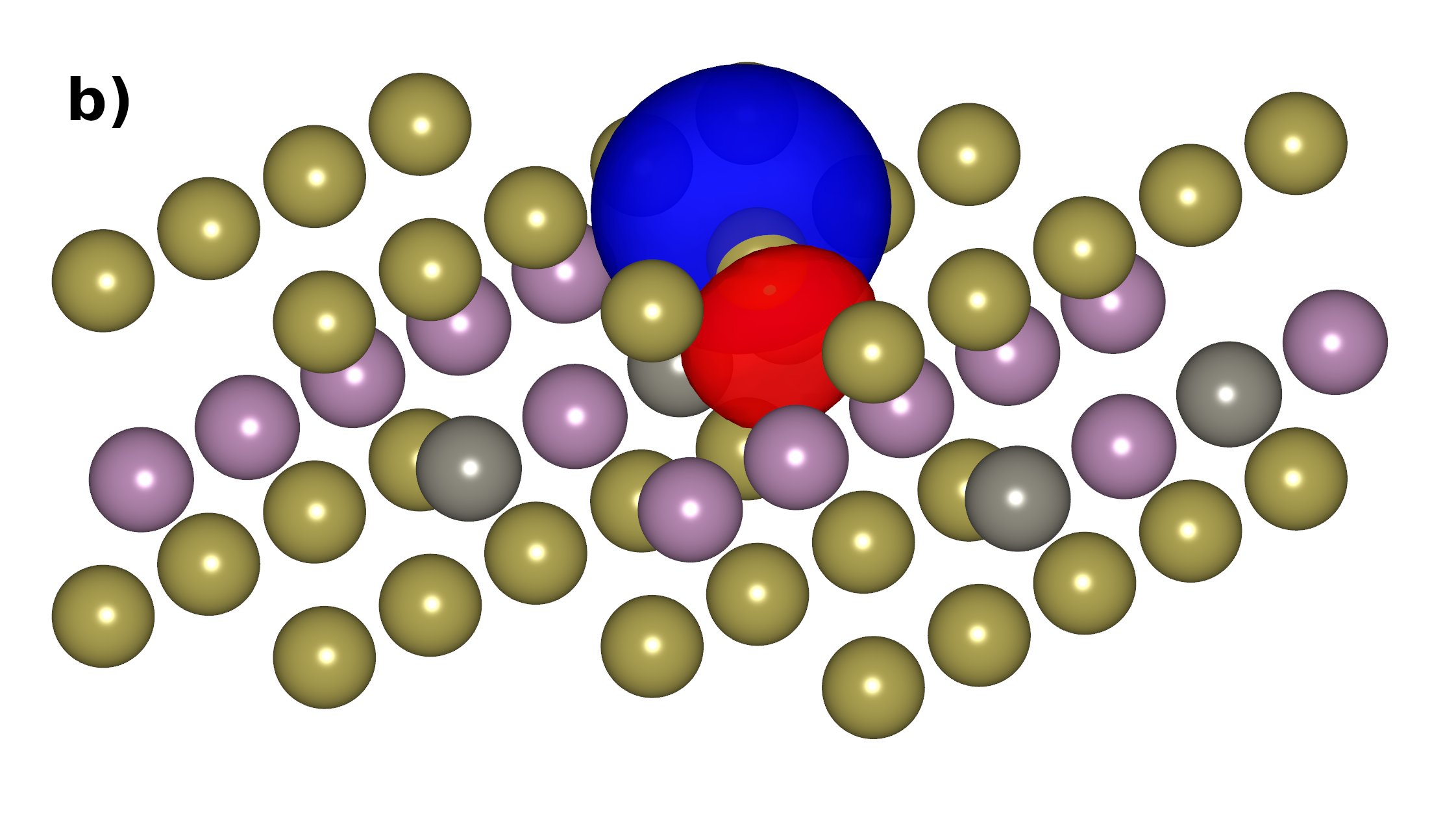}
    \includegraphics[width=\columnwidth]{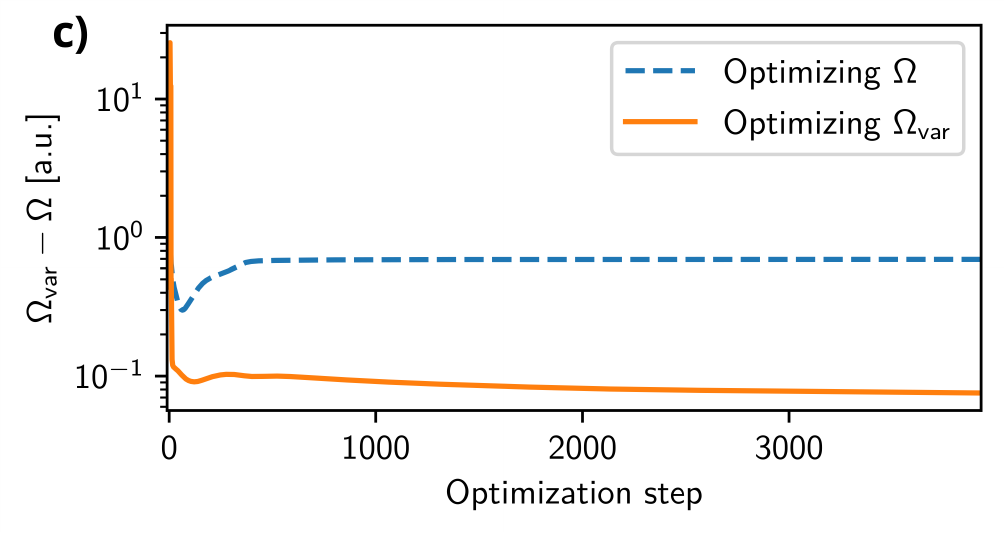}
    \caption{
        Isosurface plots of the most delocalized WF of monolayer WMo$_3$Te$_8$ obtained with (a) the standard localization functional $\Omega$ yielding a spread of $s_{\mathrm{max}} = 21.5$ \si{\angstrom^2} and (b) the variance reducing functional $\mathrm{\Omega_{var}}$ yielding a spread of $s_{\mathrm{max}} = 5.1$ \si{\angstrom^2}. The Wannierisation has been performed for the minimal number of WFs consistent with an energy threshold of $E_0 = \mathrm{CBM} + 2$ \si{\electronvolt}. The isosurface level is \SI{0.7}{\angstrom^{-3/2}}. 
        (c) The value of the variance penalty term in $\mathrm{\Omega_{var}}$ (see Eq. (\ref{eq:Omega_var})) during the iterative optimization. Additional optimization steps with $\mathrm{\Omega_{var}}$ can further decrease the variance term by a factor of 2.
    }
    \label{fig:wf_vesta}
\end{figure}

For monolayer WMo$_3$Te$_8$ we obtain an average spread (see App. \ref{app:spread}) of $\bar{s} = 2.7$ \si{\angstrom^2} and a maximum spread of $s_{\mathrm{max}} = 21.5$ \si{\angstrom^2} with the standard functional $\Omega$, while the variance minimizing functional $\mathrm{\Omega_{var}}$ converges to $\bar{s} = 2.8$ \si{\angstrom^2} and $s_{\mathrm{max}} = 5.1$ \si{\angstrom^2}.
The most delocalized WFs for both functionals are shown in Figure \ref{fig:wf_vesta}.
For the band structure interpolation error (see App. \ref{app:bsdist}) we obtain $\eta = 21$ \si{\milli\electronvolt} ($\eta_{\mathrm{max}} = 420$ \si{\milli\electronvolt}) for the standard MLWFs produced by $\Omega$ and $\eta = 9$ \si{\milli\electronvolt} ($\eta_{\mathrm{max}} = 143$ \si{\milli\electronvolt}) for the spread balanced WFs generated with $\mathrm{\Omega_{var}}$.

As expected, the average localization of the WFs generated with the two different spread functionals is almost identical while the spread of the most delocalized WFs is greatly improved by $\mathrm{\Omega_{var}}$. Moreover, the variance reducing functional leads to a significant improvement in the tight-binding interpolation of the band structure. We stress that the significant improvements found with $\mathrm{\Omega_{var}}$ for this specific material may not be fully representative. Although we do generally find a significant improvement, i.e. reduction, of $s_{\mathrm{max}}$, the band interpolation error is often similar to that obtained with $\Omega$ (see Sec. \ref{sec:HT}). 

The effect of the variance penalty term on the optimization of $\mathrm{\Omega_{var}}$ is plotted in Figure \ref{fig:wf_vesta} (c).
Comparing the value of the variance term between the two iterative optimizations, with $\Omega$ and $\mathrm{\Omega_{var}}$, it is clear that the penalty term leads to a different convergence path.
For the vast majority of the materials we tested, the number of steps required for the optimization of $\mathrm{\Omega_{var}}$ was comparable to or slightly larger than required for $\Omega$.
However, due to the additional terms in the gradient of $\mathrm{\Omega_{var}}$, each step is roughly twice as expensive to evaluate in terms computational time.

\subsection{Spontaneous polarization}

The spontaneous polarization of ferrolectrics comprises a prominent example of a physical quantity that is easily accessible from Wannier functions. As shown by King-Smith and Vanderbilt \cite{King-Smith1993} the change in polarization under an adiabatic deformation can be calculated by a Berry-phase type formula. Typically the polarization of ferroelectrics is measured with respect to a centrosymmetric phase which is known to have vanishing polarization and the spontaneous polarization can then be computed by a single calculation in the polar phase.

To unravel the relation to Wannier functions it is straightforward to show that the Wannier charge centers can be written as \cite{mlwf-review}
\begin{align}
    \mathbf{r}_n\equiv\langle w_n|\mathbf{\hat r}|w_n\rangle=\frac{V}{(2\pi)^3}\int d\mathbf{k}\langle u_{n\mathbf{k}}|i\mathbf{\nabla}_\mathbf{k}|u_{n\mathbf{k}}\rangle,
\end{align}
where $V$ is the unit cell volume and $u_{n\mathbf{k}}$ is the periodic part of a Bloch function. Except for the factor of $V$ this expression is exactly the Berry phase formula for the electric contribution to the polarization for a single band and the full polarization can thus be written as
\begin{align}\label{eq:polarization}
    \mathbf{P}=\frac{1}{V}\sum_aZ_a\mathbf{r}_a-\frac{1}{V}\sum_{n\in occ}\mathbf{r}_n,
\end{align}
where $\mathbf{r}_a$ is the position of nucleus $a$ with charge $Z_a$. Eq. \eqref{eq:polarization} is formally equivalent to the Berry phase expression for the polarization and the $2\pi$ ambiguity in the Berry phase is reflected by the fact that the nuclei positions as well as the Wannier functions can be chosen in an arbitrary unit cell. This expression for the polarization has the advantage of providing a clear physical interpretation of the polarization as the dipole resulting from a diplacement of Wannier charge centers from the nuclei positions. Moreover, it strongly facilitates a microscopic analysis of how the polarization is affected by impurities or interfaces since one may monitor the shift in Wannier charge centers under an applied perturbation.

We have compared the calculated spontaneous polarization of tetragonal BaTiO$_3$ obtained with a direct implementation of the Berry phase method \cite{Gjerding2021} with that obtained from Eq. \eqref{eq:polarization} using the $\mathrm{\Omega_{var}}$ spread functional Eq. \eqref{eq:Omega_var} to construct Wannier functions from the occupied states. In the present case we have performed a full relaxation with the PBE functional using a $8\times8\times8$ $k$-point mesh and 800 eV plane wave cutoff. The result from both calculations is 45.4 $\mu \textrm{C}/\textrm{cm}^2$, which is in agreement with previous calculations \cite{Castelli2019, Petralanda2020}. As expected, the new type of spread balanced Wannier functions thus reproduce the result obtained with standard maximally localized Wannier functions.

\subsection{Complex systems}

\begin{table*}[t]
\begin{tabular}{l@{\hspace{1em}}c@{\hspace{1em}}c@{\hspace{1em}}c@{\hspace{1em}}c@{\hspace{1em}}c}
  \toprule
   & & \multicolumn{2}{c}{$\Omega$} & \multicolumn{2}{c}{$\mathrm{\Omega_{var}}$} \\
  \cmidrule(rr){3-4} \cmidrule(rr){5-6}
  System & $N_w$ & $\langle s_{\mathrm{max}}\rangle$ [\si{\angstrom^2}] & $\eta_{\mathrm{max}}$ [\si{\milli\electronvolt}] & $\langle s_{\mathrm{max}}\rangle$ [\si{\angstrom^2}] & $\eta_{\mathrm{max}}$ [\si{\milli\electronvolt}] \\ [2pt]
  \midrule 
  NV center in diamond & 127 & \num{6.8 \pm 0.3} & \num{8} & \num{4.46 \pm 0.04} & \num{8} \\ [2pt]
  Adsorbed H on Ru slab & 133 & \num{8 \pm 1} & \num{158} & \num{4.6 \pm 0.5} & \num{172} \\ [2pt]
  Adsorbed N on Ru slab & 134 & \num{6.8 \pm 0.7} & \num{89} & \num{4.6 \pm 0.6} & \num{82} \\ [2pt]
  Adsorbed O on Ru slab & 135 & \num{6.9 \pm 0.5} & \num{77} & \num{4.7 \pm 0.4} & \num{117} \\ [2pt]
  \bottomrule
\end{tabular}
\caption{
    Comparison of the spread of the most delocalized WF and the maximum band interpolation error obtained for WFs generated with the standard spread functional $\Omega$ and the variance reducing functional $\mathrm{\Omega_{var}}$ for the NV center in diamond and H, N, O adsorbed on a Ru(111) surface slab.
    The minimal $N_w$ was used for all materials.
    $\langle s_{\mathrm{max}}\rangle$ is the average over 5 calculations with different random seeds (standard deviation also shown). 
    $\eta_{\mathrm{max}}$ is the maximum band interpolation error for the most localized set of WFs (lowest maximum spread) obtained in the 5 calculations with different random seeds.
}
\label{tab:complex}
\end{table*}

While the generation of well localized Wannier functions is usually relatively straightforward for simple systems (small number of atoms, isolated group of bands, etc.), it can be significantly more challenging in the general case. To test the variance reducing localization functional on more complex systems we compare its performance to the standard $\Omega$ functional for a nitrogen-vacancy (NV) defect center in a diamond crystal and a Ru(111) surface slab with adsorbed H, N, and O atoms, respectively.
The results are summarized in Table \ref{tab:complex} and confirm the previous conclusions. In particular, the spread of the most delocalized WF is significantly reduced when using the $\mathrm{\Omega_{var}}$ functional.

Additional computational details are provided in App. \ref{app:complex}. 

\subsection{Towards high-throughput applications}
\label{sec:HT}

\begin{figure*}[p!]
    \centering
    \includegraphics[width=\columnwidth]{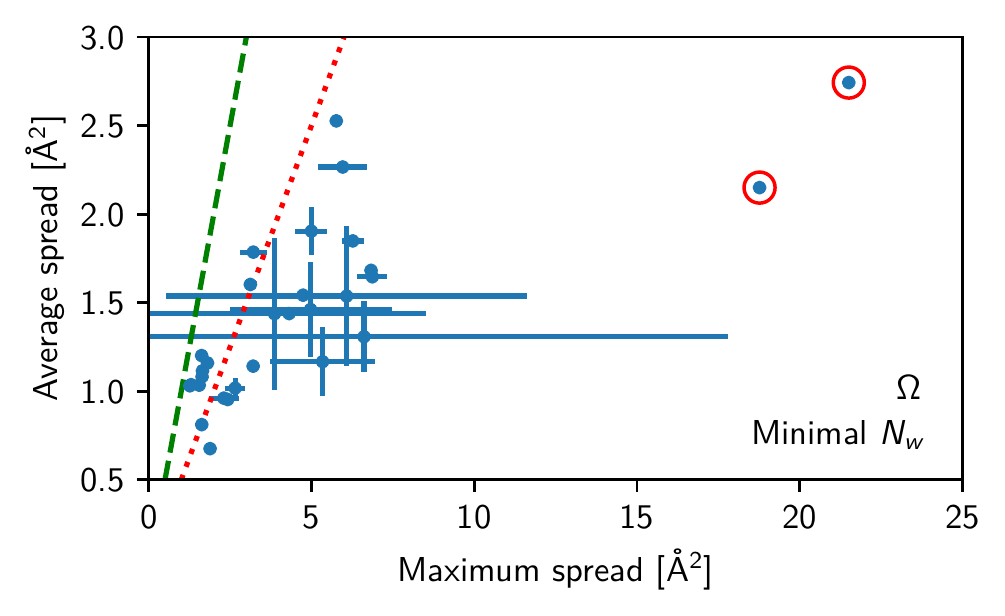} \hfill
    \includegraphics[width=\columnwidth]{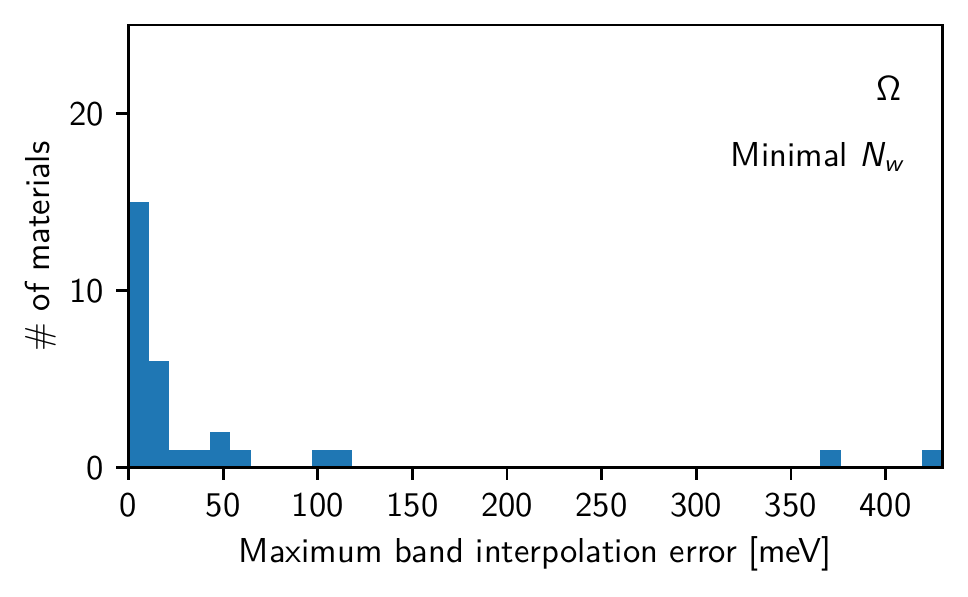}
    \includegraphics[width=\columnwidth]{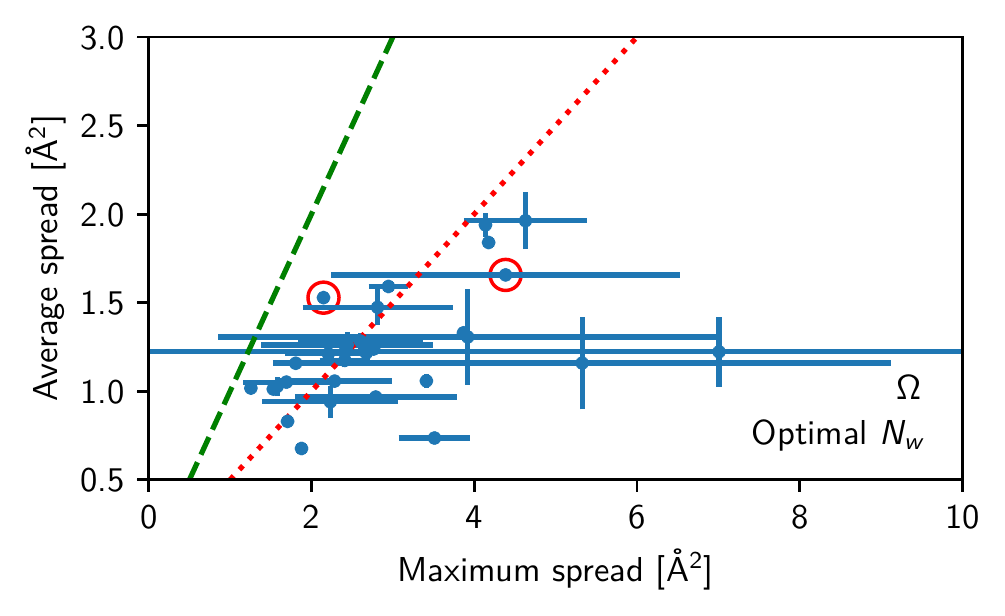} \hfill
    \includegraphics[width=\columnwidth]{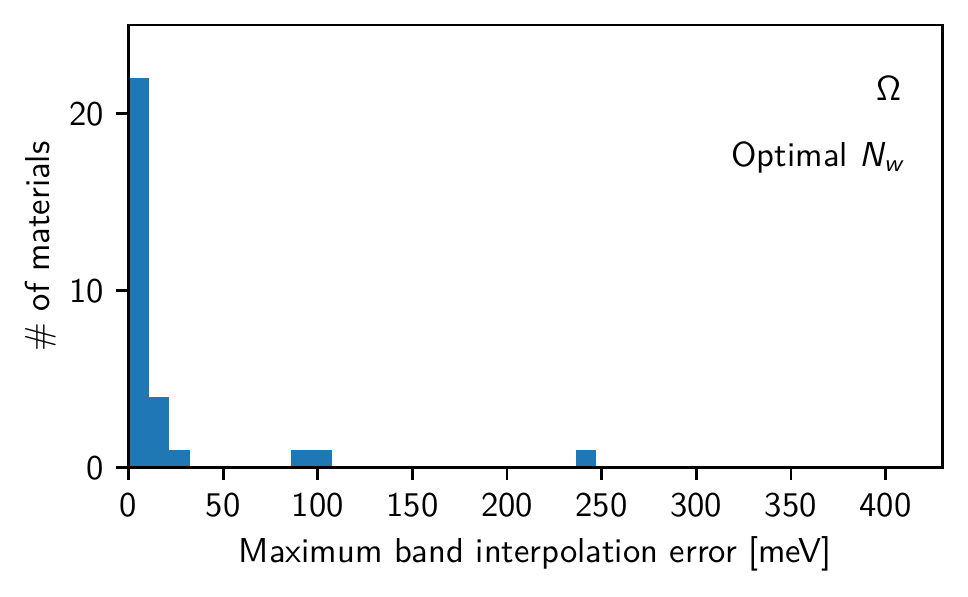}
    \includegraphics[width=\columnwidth]{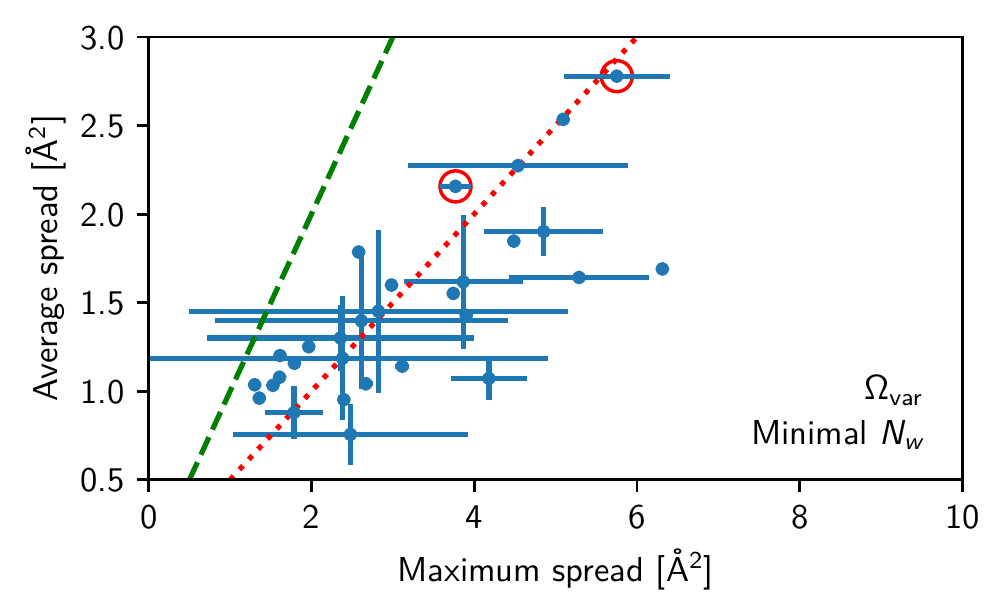} \hfill
    \includegraphics[width=\columnwidth]{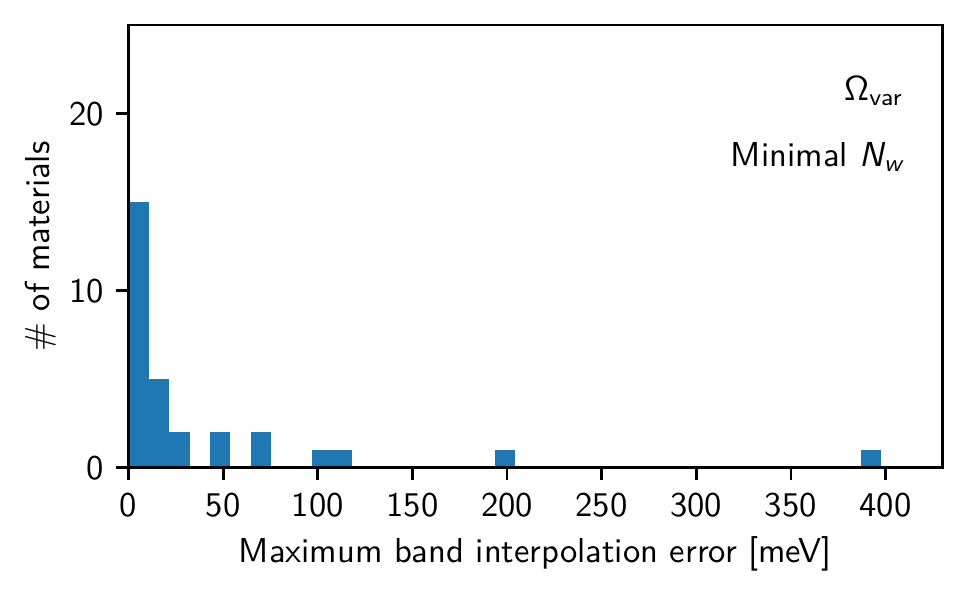}
    \includegraphics[width=\columnwidth]{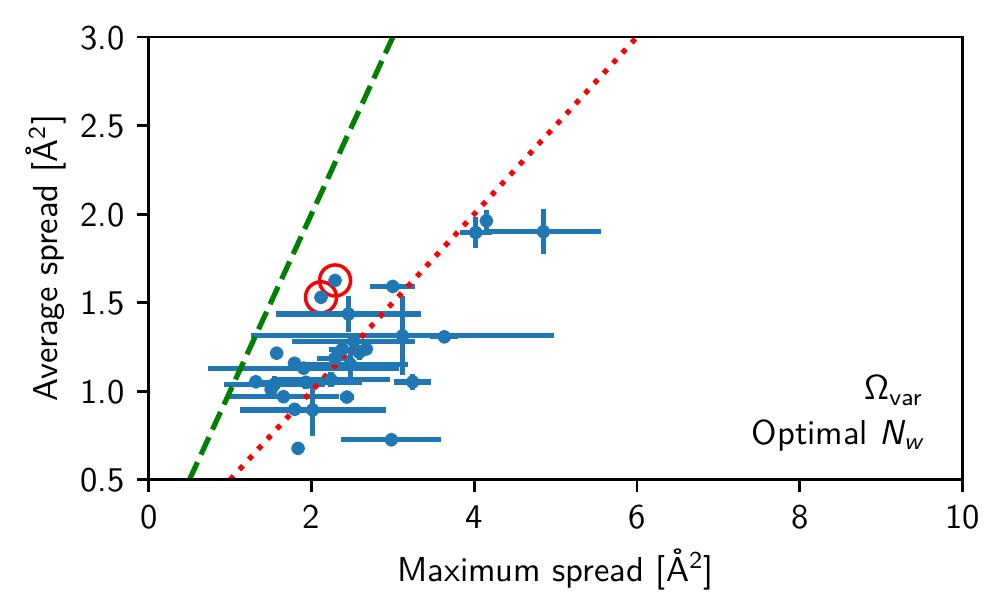} \hfill
    \includegraphics[width=\columnwidth]{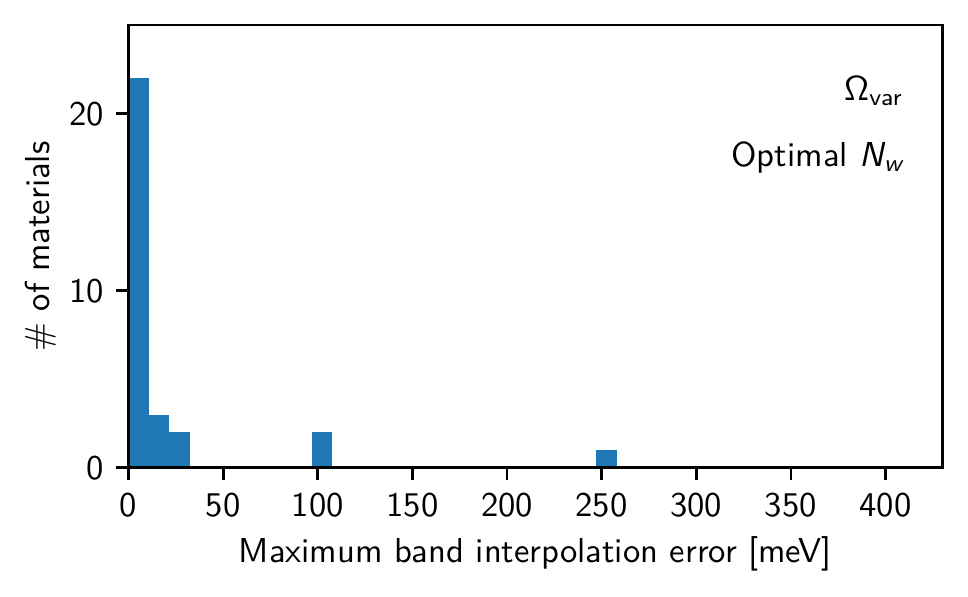}
    \caption{
        Results for the set of 2D materials with different choices for $N_w$ and the localization functional.
        The energy threshold is set at CBM ($E_F$) + \SI{2}{\electronvolt} for insulators (metals).
        Left column: Spread of WFs for all the materials in the test set averaged over 5 runs with different initializations. The standard deviation is indicated by lines.
        The dashed green line is $s_{\mathrm{max}} = \bar{s}$, the dotted red line is $s_{\mathrm{max}} = 2 \bar{s}$.
        The red circles indicate results for WMo$_3$Te$_8$ and ZrTi$_3$Te$_8$.
        Right column: The maximum band interpolation error.
        Results from the most localized set of WFs, out of the 5 calculations with different random seeds.
    }
    \label{fig:test_set}
\end{figure*}

In the previous sections we have demonstrated our new Wannierisation scheme for several different types of systems. In this section, we apply the scheme to a larger set of materials comprising 30 atomically thin two-dimensional (2D) materials randomly selected from the Computational 2D Materials Database \cite{c2db}. The set includes 22 materials with finite band gaps and 8 metals, and covers several different crystal lattices and a large set of chemical elements.
The complete list of materials is provided in App. \ref{app:materials} and the computational details in App. \ref{app:methods}. Although we focus on 2D materials we expect our results to be representative for general material types. 

We applied our Wannierisation scheme to each material and determined the optimal $N_w$ as the one yielding the lowest maximum spread, $s_{\mathrm{max}}$, see Sec. \ref{sec:selecting}. For each value of $N_w$ in the range from $N_w^{\mathrm{min}}$ to $N_w^{\mathrm{min}}+5$, we performed 5 independent Wannierisations using different initializations (differing due to the arbitrariness in the position of the $s$-orbitals). Following the procedure visualised in Fig. \ref{fig:gaas-best}(top), we used the average $s_{\mathrm{max}}$ over the 5 sets of WFs to determine the optimal $N_w$. In all cases, we include eigenstates up to \SI{2}{\electronvolt} above the conduction band minimum (Fermi level) for insulating (metallic) materials.

The results of the Wannierisation procedure for the 30 materials are summarised in Fig. \ref{fig:test_set}. The left column shows the average and maximal spreads of the WFs generated with the $\Omega$ and $\Omega_{\mathrm{var}}$ functionals for the minimal and optimal number of WFs, respectively. Note the different scales on the axes. The symbols indicate the mean values of $\bar{s}$ and $s_{\mathrm{max}}$ over the 5 initializations, and the lines indicate the standard deviation. The green and red dashed lines indicate $\langle \bar{s} \rangle = \langle s_{\mathrm{max}}\rangle$ and  $\langle \bar{s} \rangle = 2\langle s_{\mathrm{max}}\rangle$, respectively.    

For generating well-localized WFs, the number of WFs, i.e. using $N_w^{opt}$ rather than $N_w^{min}$, was found to be more critical than the type of spread functional, i.e. $\Omega$ versus $\Omega_{\mathrm{var}}$. However, the use of $\Omega_{\mathrm{var}}$ instead of $\Omega$ does lead to a significant improvement in the localisation for a fixed $N_w$. This improvement is most pronounced for non-optimal values of $N_w$, e.g. the minimal $N_w$. In particular, for the two materials WMo$_3$Te$_8$ and ZrTi$_3$Te$_8$ (both with $N_w>60$ and indicated by red circles), we were not able to localize all the WFs when using the $\Omega$-functional and the minimal $N_w$. This issue did not occur with the $\Omega_{\mathrm{var}}$-functional. Even when using the optimal $N_w$, the improvement by the $\Omega_{\mathrm{var}}$-functional is significant. Not only do we obtain better localisation of the least localised WF ($s_{\mathrm{max}}$) without sacrificing the average localisation ($\bar{s}$), the standard deviations on both $\bar{s}$ and $s_{\mathrm{max}}$ is also lowered.

The band interpolation errors (again averaged over the 5 initializations) are shown in the right column of Fig. \ref{fig:test_set}. As previously found and discussed in Section \ref{sec:optimal}, the errors decrease significantly when using the optimized $N_w$ as compared to the minimal $N_w$. With the optimal $N_w$ the majority of the materials show a maximum error below \SI{20}{\milli\electronvolt}. A few materials show higher band errors, which is related to significant band crossings (band entanglement) with higher energy bands close to the $E_0$ energy cutoff. We note in passing that if accurate band interpolation in this region is required, one may simply increase $E_0$, which will push the inaccuracies to higher band energies. 

While the $\Omega_{\mathrm{var}}$ spread functional improves the localisation properties of WFs, it does not represent a significant improvement over $\Omega$ in terms of the band interpolation error. This shows that the band interpolation error is not directly correlated with the localisation properties of the WFs. More precisely, it is not directly correlated with the spread of the most delocalised WF, $s_{\mathrm{max}}$, which is always significantly and consistently reduced by using $\Omega_{\mathrm{var}}$. We stress, however, that there are many other applications of WFs where robust localisation of all WFs, is of key importance. These include the interpolation of electron-phonon matrix elements\cite{giustino2007electron,sjakste2015wannier}, calculation of Berry curvatures and conductivities\cite{wang2006ab}, and basis sets for electron transport calculations based on non-equilibrium Green's functions\cite{calzolari2004ab,strange2008benchmark}.   

\begin{table*}[t]
\begin{tabular}{l@{\hspace{1em}}c@{\hspace{1em}}c@{\hspace{1em}}c@{\hspace{1em}}c@{\hspace{1em}}c}
  \toprule
  Functional & $N_w$ & $\langle \langle \bar{s} \rangle_{\mathrm{init}} \rangle_{\mathrm{mat}}$ [\si{\angstrom^2}] & $\langle \sigma_{\mathrm{init}}(\bar{s}) \rangle_{\mathrm{mat}}$ [\si{\angstrom^2}] & $\langle \langle s_{\mathrm{max}} \rangle_{\mathrm{init}} \rangle_{\mathrm{mat}}$ [\si{\angstrom^2}] & $\langle \sigma_{\mathrm{init}}(s_{\mathrm{max}}) \rangle_{\mathrm{mat}}$ [\si{\angstrom^2}] \\ [2pt]
  \midrule 
  $\mathrm{\Omega}$ & Minimal & \num{1.44 \pm 0.09} & \num{0.06 \pm 0.02} & \num{4.8 \pm 0.8} & \num{1.0 \pm 0.4} \\ [2pt]
  $\mathrm{\Omega}$ & Optimal & \num{1.24 \pm 0.06} & \num{0.05 \pm 0.01} & \num{2.9 \pm 0.2} & \num{1.0 \pm 0.4} \\ [2pt]
  $\mathrm{\Omega_{var}}$ & Minimal & \num{1.46 \pm 0.09} & \num{0.08 \pm 0.02} & \num{3.2 \pm 0.3} & \num{0.5 \pm 0.1} \\ [2pt]
  $\mathrm{\Omega_{var}}$ & Optimal & \num{1.23 \pm 0.06} & \num{0.05 \pm 0.01} & \num{2.5 \pm 0.1} & \num{0.4 \pm 0.1} \\ [2pt]
  \bottomrule
\end{tabular}
\caption{
    Comparison of the spread distribution over the entire set of 2D materials.
    The definitions of $\bar{s}_{\mathrm{tot}}$, $\bar{\sigma}_{\mathrm{tot}}$, $\bar{s}_{\mathrm{max}}$ and $\bar{\sigma}_{\mathrm{max}}$ are given in Eqs. \ref{eq:s_tot}, \ref{eq:sigma_tot} and the following text.
    For each value we also report the standard deviation of the mean (i.e. the standard error).
}
\label{tab:stats}
\end{table*}

The effect of penalizing the spread variance in $\Omega_{\mathrm{var}}$ can be illustrated by considering the localisation properties of the resulting sets of WFs averaged over all 30 materials for a fixed $N_w$.
Specifically, we fix $N_w$ at either its minimal or optimal value, and consider the mean of the average spread
\begin{equation}
\label{eq:s_tot}
    \langle \langle \bar{s} \rangle_{\mathrm{init}} \rangle_{\mathrm{mat}} = \frac{1}{N_{\mathrm{mat}}} \sum^{N_{\mathrm{mat}}}_{m=1} \left( \frac{1}{N_{\mathrm{init}}} \sum^{N_{\mathrm{init}}}_{i=1} \bar{s}_i^m \right)
\end{equation}
and the mean standard deviation of the average spread over 5 initialisations
\begin{equation}
\label{eq:sigma_tot}
    \langle \sigma_{\mathrm{init}}(\bar{s}) \rangle_{\mathrm{mat}} = \frac{1}{N_{\mathrm{mat}}} \sum^{N_{\mathrm{mat}}}_{m=1}  \sqrt{\mathrm{Var}_{\textrm{init}}[\bar{s}_i^m]},
\end{equation}
where $N_{\mathrm{mat}}$ is the total number of materials (here $N_{\mathrm{mat}}=30$), $N_{\mathrm{init}}$ is the number of independent optimizations with different initial guess (here $N_{\mathrm{init}}=5$), $\bar{s}_i^m$ is the average spread of the WFs of material $m$ with initialization $i$, and $\mathrm{Var}_{\mathrm{init}}$ is the variance of the $N_{\mathrm{init}}$ independent runs.
In the same way, we can define $\langle \langle s_{\mathrm{max}} \rangle_{\mathrm{init}} \rangle_{\mathrm{mat}}$ and $\langle \sigma_{\mathrm{init}}(s_{\mathrm{max}}) \rangle_{\mathrm{mat}}$ by replacing $\bar{s}$ with $s_{\mathrm{max}}$ in the above equations. 
Note, that $\langle \sigma_{\mathrm{init}}(\bar{s}) \rangle_{\mathrm{mat}}$ and $\langle \sigma_{\mathrm{init}}(s_{\mathrm{max}}) \rangle_{\mathrm{mat}}$ measure the variation in the average and maximal spread in dependence of the initial guess, i.e. the \emph{robustness} of the Wannierisation. In particular, it does not reflect the variation of the spreads within a given set of WFs. As demonstrated on several places in the paper, the latter is always significantly and consistently reduced when using $\Omega_{\mathrm{var}}$. 

Table \ref{tab:stats} shows the results for the four quantities $\langle \langle \bar{s} \rangle_{\mathrm{init}} \rangle_{\mathrm{mat}}$, $\langle \langle s_{\mathrm{max}} \rangle_{\mathrm{init}} \rangle_{\mathrm{mat}}$, $\langle \sigma_{\mathrm{init}}(\bar{s}) \rangle_{\mathrm{mat}}$, and $\langle \sigma_{\mathrm{init}}(s_{\mathrm{max}}) \rangle_{\mathrm{mat}}$ for each of the spread functionals $\Omega$ and $\Omega_{\mathrm{var}}$.
These numbers summarise the information in Fig. \ref{fig:test_set}. On this basis we conclude that the penalization of the spread variance as done in the $\mathrm{\Omega_{var}}$-functional, does not affect the average spread of the WFs, but consistently decreases the maximum spread and improves the robustness with respect to the choice of initial orbitals.

\subsection{Data availability}

The data presented in this article together with information required for its reproduction, are available in an online open access archive \cite{data_repo}. For more details see also App. \ref{app:software}.

\section{Conclusions}

We have introduced a new localization functional for generating Wannier functions (WFs) with balanced spread distributions.  By penalizing the variance of the spread distribution, the algorithm becomes less prone to produce individual WFs with large spreads. It thereby resolves a well known problem of standard Wannierisation schemes, which is particularly important when applied to complex systems. Furthermore, we have proposed a general and fully automatic algorithm for selecting the optimal number of WFs and the initial set of orbitals for the Wannierisation procedure. Application to an extensive test suite comprising both bulk and monolayer materials, point defects and atoms on metal slabs, consistently show that these algorithms comprise a highly robust approach for generating WFs, which should be particularly useful for applications to complex systems and/or high-throughput studies. The methods are implemented in Python and are available as part of the open-source Atomic Simulation Environment (ASE) \cite{ase}.  

\section{Acknowledgments}
The Center for Nanostructured Graphene (CNG) is sponsored by the Danish National Research Foundation, Project DNRF103. This project has received funding from the European Research Council (ERC) under the European Union’s Horizon 2020 research and innovation program grant agreement No 773122 (LIMA). K. S. T. is a Villum Investigator supported by VILLUM FONDEN (grant no. 37789).

\appendix

\section{Wannier function spread}
\label{app:spread}

We define
\begin{equation}
  s_n = - \frac{1}{(2 \pi)^2} \sum_{\alpha=1}^{N_G} W_{\alpha} \log \left(|Z_{\alpha, nn}|^2 \right).
\end{equation}
This quantity is an approximation for the spread of a single WF in \si{\angstrom^2} \cite{berghold}, if the weights $W_{\alpha}$ retain the physical dimensions.
This definition has the feature of exponentially increasing in case of completely delocalized WFs, which can help with the identification of localization issues but at the same time it does not retain any physical meaning for delocalized WFs.
We note that most definitions of spread or localization functional have a convergence with the $k$-point grid density, as observed and studied in Ref. \cite{stengel}.
In the present work we often refer to $\bar{s}$, the average of $s_n$ over the set of WFs, and $s_{max}$, the maximum value of $s_n$ over the set of WFs.

\section{Band interpolation error}
\label{app:bsdist}

In order to quantitatively estimate the quality of the band interpolation we introduce a quantity called band interpolation error, that measures the difference between the Kohn-Sham and the Wannier-interpolated band structures.
The definition is the same that appears Refs. \cite{auto-wan, variational_wan} and allows to easily compare the results.
The band interpolation error is computed up to the energy threshold $E_0$, introduced in Section \ref{sec:powf}.

If the band structure has a gap at the energy threshold, then the interpolation error is computed only on the energy bands below the gap.
We define the average band interpolation error $\eta$ and the maximum contribution to the band interpolation error $\eta_{\mathrm{max}}$ as
\begin{equation}
  \eta = \sqrt{\sum_{n\vb{k}}\frac{(\varepsilon^{\mathrm{KS}}_{n\vb{k}} - \varepsilon^{\mathrm{Wan}}_{n\vb{k}})^2}{N_b N_{\vb{k}}}}
\end{equation}
\begin{equation}
  \eta_{\mathrm{max}} = \text{max}_{n\vb{k}}(|\varepsilon^{\mathrm{KS}}_{n\vb{k}} - \varepsilon^{\mathrm{Wan}}_{n\vb{k}}|).
\end{equation}
where $\varepsilon^{KS}$ and $\varepsilon^{Wan}$ are the Kohn-Sham eigenvalues and their Wannier interpolation respectively, $N_b$ is the number of bands and $N_{\vb{k}}$ the number of $k$ points.

In the case of entangled bands we introduce a different definition which uses a smearing function to weight the contributions from the bands around the energy threshold
\begin{equation}
  \eta = \sqrt{\frac{\sum_{n\vb{k}}(\varepsilon^{\mathrm{KS}}_{n\vb{k}} - \varepsilon^{\mathrm{Wan}}_{n\vb{k}})^2 \tilde{f}_{n\vb{k}}}{\sum_{n\vb{k}} \tilde{f}_{n\vb{k}}}}
\end{equation}
\begin{equation}
  \eta_{\mathrm{max}} = \text{max}_{n\vb{k}}(\tilde{f}_{n\vb{k}} |\varepsilon^{\mathrm{KS}}_{n\vb{k}} - \varepsilon^{\mathrm{Wan}}_{n\vb{k}}|)
\end{equation}
where $\tilde{f}_{n\vb{k}} = \sqrt{f^{\mathrm{KS}}_{n\vb{k}}(\nu, \tau) f^{\mathrm{Wan}}_{n\vb{k}}(\nu, \tau)}$ and $f_{n\vb{k}}(\nu, \tau)$ is the Fermi-Dirac distribution for the state at energy $\varepsilon_{n\vb{k}}$, $\nu$ is a fictitious chemical potential fixed at $E_0$ and $\tau$ is a smearing width fixed to \SI{0.1}{\electronvolt}.

\section{Computational details}
\label{app:methods}

For each material in this work we performed a self-consistent PBE calculation with the GPAW code\cite{gpaw} using a Monkhorst-Pack grid \cite{mp-grid} with a minimum density of 5 $k$-points per \si{\per\angstrom} and a real-space grid with a spacing of \SI{0.2}{\angstrom}. Several unoccupied states were included in the calculation.
The band structure calculation, used to evaluate the band interpolation accuracy, was performed by fixing the density to the value from the previous self-consistent calculation and evaluating the band structure along a path with minimum density of 50 $k$-points per \si{\per\angstrom}.
Finally, we computed the WFs with ASE \cite{ase}, starting from the self-consistent DFT Bloch states.
For all materials in Sec. \ref{sec:HT} we used the structure available in the public C2DB database \cite{c2db}.

\section{Methods for complex systems}
\label{app:complex}

For the NV center in diamond we substituted one carbon atom with a nitrogen atom in a 64 atom cubic unit cell of bulk diamond.
We then proceeded in the structure relaxation using GPAW \cite{gpaw}, with a 2x2x2 Monkhorst-Pack grid of $k$-points in the Brillouin zone (BZ), a plane waves basis set with an energy cutoff of \SI{400}{\electronvolt} and an initial charge of \num{-1}.
The structure was optimized with the LBFGS algorithm \cite{lbfgs} and the force threshold was set to \SI{0.05}{\electronvolt\per\angstrom}.
After the structure was relaxed we ran a self-consistent and a non self-consistent calculation, with a BZ sampling density of \num{2} and \num{5} $k$-points per \si{\per\angstrom} respectively.
The self-consistent calculation was converged for every state up to \SI{4}{\electronvolt} from the conduction band minimum (CBM).
The rest of the methods were in line with Section \ref{sec:HT}.

With the adsorption systems we followed a similar workflow with ASE \cite{ase}. 
We created a 2x2x4 slab of ruthenium with HCP(0001) surface and a vacuum layer of \SI{7}{\angstrom} along the $z$ direction.
Then the adsorbate was placed on the HCP site at \SI{1}{\angstrom}, \SI{1.108}{\angstrom}, \SI{1.257}{\angstrom} for H, N, O respectively.
The structure was then relaxed, fixing the first two layers of Ru, using the finite-difference method and the PBE \cite{pbe} xc-functional in GPAW, again using LBFGS as optimization algorithm and \SI{0.05}{\electronvolt\per\angstrom} as forces threshold.
After the relaxation, a self-consistent and a non self-consistent calculations were ran on each system, following the workflow for the NV center in diamond.

\section{List of 2D materials}
\label{app:materials}

The 2D structures in Table \ref{tab:materials} are picked from the C2DB database \cite{c2db}.
The values for the maximum spread are the mean and the standard deviation over 5 calculations with different random seeds for the initial guess.
The standard deviation for the maximum band interpolation error is computed with the same method.
For the maximum band interpolation and the number of WFs (the latter varies only in the case of optimal $N_w$) we use the value from the most localized set (lowest maximum spread) over 5 calculations, instead of the mean value.
In the last row we show the average value and the average standard deviation for each column.

\begin{table*}
\begin{tabular}{l@{\hspace{1em}}l@{\hspace{1em}}r@{\hspace{1em}}r@{\hspace{1em}}r@{\hspace{1em}}r@{\hspace{1em}}r@{\hspace{1em}}r}
  \toprule
   & & \multicolumn{3}{c}{Minimal $N_w$ / $\Omega$} & \multicolumn{3}{c}{Optimal $N_w$ / $\mathrm{\Omega_{var}}$} \\
  \cmidrule(rr){3-5} \cmidrule(rr){6-8}
  Formula    & Crystal type & $N_w$ & $\langle s_{\mathrm{max}}\rangle$ [\si{\angstrom^2}] & $\eta_{\mathrm{max}}$ [\si{\milli\electronvolt}] & $N_w$ & $\langle s_{\mathrm{max}}\rangle$ [\si{\angstrom^2}] & $\eta_{\mathrm{max}}$ [\si{\milli\electronvolt}] \\ [2pt]
  \midrule

  Ag$_2$Br$_2$       & \texttt{AB-129-bc} & 26 & \num{2.3 \pm 0.5} & \num{11.7 \pm 0.2} & 26 & \num{1.7 \pm 0.7} & \num{11.6 \pm 0.1} \\ [4pt]
  Al$_2$Cl$_2$O$_2$  & \texttt{ABC-59-ab} & 20 & \num{4.3 \pm 0.2} & \num{13 \pm 3} & 23 & \num{3.6 \pm 0.2} & \num{5.0 \pm 0.7} \\ [4pt]
  Au$_2$S$_2$        & \texttt{AB-10-fgm} & 19 & \num{3.21 \pm 0.08} & \num{24 \pm 5} & 22 & \num{2.6 \pm 0.1} & \num{5 \pm 3} \\ [4pt]
  BrClTi             & \texttt{ABC-59-ab} & 18 & \num{2.43 \pm 0.01} & \num{5.1 \pm 0.1} & 18 & \num{2.44 \pm 0.09} & \num{5.2 \pm 0.3} \\ [4pt]
  Br$_2$Hf$_2$S$_2$  & \texttt{ABC-156-ac} & 29 & \num{6.3 \pm 0.3} & \num{373 \pm 6} & 32 & \num{1.57 \pm 0.07} & \num{105.3 \pm 0.4} \\ [4pt]
  C$_2$H$_2$         & \texttt{AB-164-d} & 8 & \num{5.0 \pm 0.5} & \num{13.3 \pm 0.4} & 8 & \num{4.9 \pm 0.7} & \num{8 \pm 2} \\ [4pt]
  CF$_2$Y$_2$        & \texttt{AB2C2-164-bd} & 23 & \num{6.9 \pm 0.5} & \num{36 \pm 1} & 27 & \num{3.2 \pm 0.2} & \num{18 \pm 4} \\ [4pt]
  CH$_2$O$_2$V$_2$   & \texttt{AB2C2D2-164-bd} & 31 & \num{5.3 \pm 1.6} & \num{111 \pm 58} & 35 & \num{3 \pm 0.6} & \num{254 \pm 75} \\ [4pt]
  Cl$_2$Sc$_2$Se$_2$ & \texttt{ABC-59-ab} & 32 & \num{1.64 \pm 0.05} & \num{6.3 \pm 0.1} & 33 & \num{1.9 \pm 0.7} & \num{6.5 \pm 0.6} \\ [4pt]
  Cr$_2$W$_2$S$_8$   & \texttt{ABC4-28-bcd} & 56 & \num{7 \pm 11} & \num{8 \pm 48} & 59 & \num{2 \pm 1} & \num{5.2 \pm 0.9} \\ [4pt]
  CSiH$_2$           & \texttt{ABC2-156-ab} & 9 & \num{5.763 \pm 0.001} & \num{6 \pm 2} & 13 & \num{4.15 \pm 0.02} & \num{3.0 \pm 0.6} \\ [4pt]
  Hf$_2$Cl$_4$       & \texttt{AB2-11-e} & 30 & \num{4.74 \pm 0.01} & \num{62 \pm 1} & 34 & \num{2.67 \pm 0.04} & \num{30 \pm 1} \\ [4pt]
  HgI$_2$            & \texttt{AB2-115-dg} & 17 & \num{1.806 \pm 0.001} & \num{2.9 \pm 0.1} & 17 & \num{1.792 \pm 0.001} & \num{2.9 \pm 0.1} \\ [4pt]
  I$_2$O$_2$Rh$_2$   & \texttt{ABC-59-ab} & 32 & \num{1.63 \pm 0.01} & \num{4 \pm 0.1} & 35 & \num{2 \pm 0.9} & \num{3 \pm 280} \\ [4pt]
  In$_2$S$_2$        & \texttt{AB-164-cd} & 22 & \num{6.83 \pm 0.01} & \num{52 \pm 2} & 26 & \num{2.4 \pm 0.2} & \num{5 \pm 2} \\ [4pt]
  Ir$_2$Br$_6$       & \texttt{AB3-162-dk} & 40 & \num{1.303 \pm 0.001} & \num{5.6 \pm 0.1} & 41 & \num{1.32 \pm 0.03} & \num{5.1 \pm 0.2} \\ [4pt]
  ISSb               & \texttt{ABC-156-abc} & 17 & \num{6 \pm 6} & \num{4.5 \pm 74} & 17 & \num{2.5 \pm 0.8} & \num{5 \pm 2} \\ [4pt]
  MoSe$_2$           & \texttt{AB2-187-bi} & 17 & \num{1.63 \pm 0.04} & \num{6.6 \pm 0.1} & 17 & \num{3 \pm 2} & \num{7 \pm 32} \\ [4pt]
  N$_2$O$_2$Zr$_3$   & \texttt{A2B2C3-187-bghi} & 37 & \num{2.7 \pm 0.3} & \num{9.5 \pm 20} & 39 & \num{1.8 \pm 0.7} & \num{9.2 \pm 0.9} \\ [4pt]
  Nb$_2$I$_4$        & \texttt{AB2-11-e} & 35 & \num{3.12 \pm 0.01} & \num{100.1 \pm 0.2} & 38 & \num{3 \pm 0.3} & \num{97 \pm 4} \\ [4pt]
  PdS$_2$            & \texttt{AB2-164-bd} & 13 & \num{1.26 \pm 0.02} & \num{2 \pm 0.1} & 13 & \num{2.5 \pm 0.7} & \num{2.0 \pm 0.6} \\ [4pt]
  PtSe$_2$           & \texttt{AB2-164-bd} & 16 & \num{1.652 \pm 0.001} & \num{13.9 \pm 0.1} & 18 & \num{2.3 \pm 0.2} & \num{7 \pm 5} \\ [4pt]
  Ru$_2$Se$_4$       & \texttt{AB2-11-e} & 34 & \num{1.55 \pm 0.01} & \num{9.2 \pm 0.4} & 34 & \num{2.2 \pm 0.7} & \num{8.0 \pm 0.5} \\ [4pt]
  ScSe$_2$           & \texttt{AB2-164-bd} & 12 & \num{4 \pm 5} & \num{8 \pm 189} & 16 & \num{1.5 \pm 0.01} & \num{1.8 \pm 0.4} \\ [4pt]
  SnTe$_2$           & \texttt{AB2-164-bd} & 14 & \num{3.2 \pm 0.4} & \num{16.6 \pm 0.1} & 17 & \num{2.5 \pm 0.9} & \num{15 \pm 3} \\ [4pt]
  SrCl$_2$           & \texttt{AB2-164-bd} & 17 & \num{6 \pm 0.8} & \num{0.8 \pm 0.3} & 21 & \num{4 \pm 0.2} & \num{1.2 \pm 0.3} \\ [4pt]
  TaS$_2$            & \texttt{AB2-187-bi} & 13 & \num{5 \pm 2} & \num{16 \pm 470} & 17 & \num{1.6 \pm 0.6} & \num{3.1 \pm 0.1} \\ [4pt]
  WMo$_3$Te$_8$      & \texttt{AB3C8-1-a} & 64 & \num{21.51 \pm 0.06} & \num{420 \pm 50} & 68 & \num{2.12 \pm 0.05} & \num{10.6 \pm 0.5} \\ [4pt]
  WO$_2$             & \texttt{AB2-187-bi} & 18 & \num{1.88 \pm 0.01} & \num{1.8 \pm 0.1} & 18 & \num{1.835 \pm 0.002} & \num{1.8 \pm 0.1} \\ [4pt]
  ZrTi$_3$Te$_8$     & \texttt{AB3C8-1-a} & 58 & \num{18.77 \pm 0.03} & \num{52 \pm 4} & 62 & \num{2.29 \pm 0.06} & \num{25 \pm 7} \\  [2pt]
  \midrule 
  Average & - & - & \num{4.8 \pm 0.8} & \num{47 \pm 31} & - & \num{2.48 \pm 0.15} & \num{22 \pm 14} \\ [2pt]
  \bottomrule
\end{tabular}
\caption{List of the 2D materials considered in Sec. \ref{sec:HT}. The "Crystal type" in the second column follows the convention of the C2DB database and stands for stoichiometry-space group-occupied Wyckoff positions.}
\label{tab:materials}
\end{table*}

\section{Software availability}
\label{app:software}

Most of the software used and developed in this project is open-source and available for free.
The density functional theory code used is GPAW 20.10 \cite{gpaw} with version 0.9.2 of the atomic setups, together with ASE 3.20 \cite{ase}. The latter also includes the Wannier module in which the new spread functional has been implemented. 
The Wannier functions are represented with Vesta \cite{vesta}, which is available for free.
The plots are produced with Matplotlib \cite{matplotlib} and the independent calculations were run in parallel with the help of GNU parallel \cite{parallel}, both of them are open-source software.

\bibliographystyle{apsrev4-2}
\bibliography{thebibliography}

\end{document}